\documentclass[useAMS,usenatbib]{mn2e}
\usepackage{graphicx}
\usepackage{txfonts}
\usepackage{natbib}
\newcommand{\rosat}{{\it ROSAT}}
\newcommand{\xmm}{{\it XMM-Newton}}

\newcommand{\chandra}{{\it Chandra}}

\newcommand{\einstein}{{\it Einstein~Observatory}}

\def\gs{\mathrel{\hbox{\rlap{\hbox{\lower4pt\hbox{$\sim$}}}\hbox{$>$}}}}
\def\ls{\mathrel{\hbox{\rlap{\hbox{\lower4pt\hbox{$\sim$}}}\hbox{$<$}}}}

\begin{document}
   \title[]{A Systematic Study of Variability in a Sample of Ultraluminous X-ray Sources}

   \author[L.M. Heil, S. Vaughan. T.P. Roberts ]{L.M. Heil$^{1}$\thanks{lmh38@star.le.ac.uk}, S. Vaughan$^{1}$, T.P. Roberts$^{2}$ \\
	$^{1}$ X-ray \& Observational Astronomy Group, Department of Physics and Astronomy, University of Leicester, Leicester, LE1 7RH. \\
	$^{2}$ Department of Physics, University of Durham, South Road, Durham DH1 3LE.
	}

   \date{Draft \today}

   \pagerange{\pageref{firstpage}--\pageref{lastpage}} \pubyear{2008}

   \maketitle
   
   \label{firstpage}

   \begin{abstract}
    	We present results from a study of short term variability in 19
    	archival observations by \xmm\ of 16 Ultraluminous X-ray Sources
    	(ULXs). Eight observations (six sources) showed intrinsic
    	variability with power spectra in the form of either a power
    	law or broken power law-like continuum and in some cases quasi-periodic oscillations (QPOs). The
    	remaining observations were used to place upper limits on the
    	strength of possible variability hidden within. Seven observations
    	(seven sources) yielded upper limits comparable to, or higher than,
    	the values measured from those observations with detectable variations. These represented the seven 		faintest sources all with $f_{x} < 3\times10^{-12}$ erg cm$^{-2}$ s$^{-1}$.
    	In contrast there are four observations (three sources) that gave upper limits 
    	significantly lower than both the values measured from the ULX
    	observations with detectable variations, and the values expected
    	by comparison with luminous Galactic black hole X-ray binaries (BHBs) and
    	Active Galactic Nuclei (AGN) in the observed frequency bandpass ($10^{-3}$ - 1 Hz).
	This is the case irrespective of whether one assumes characteristic frequencies appropriate
    	for a stellar mass (10 M$_{\odot}$) or an intermediate mass (1000 M$_{\odot}$) black hole,
	and means that in some ULXs the variability is significantly suppressed compared to bright BHBs and AGN. 
	We discuss ways to account for this unusual suppression in terms of both observational and intrinsic effects 		and whether these solutions are supported by our results.
   \end{abstract}

 

\section{Introduction}
\label{sect:intro}

Ultraluminous X-ray sources (ULXs), point-like sources with
luminosities greater than $10^{39}$ erg s$^{-1}$, were first identified
by the \einstein\ \citep{Fabbiano89}. Initial studies of short term variability
indicated that they were likely to be accreting compact objects
\citep{Okada98}. Both spectral analysis and
long and short term variability studies suggested similarities between
ULXs and Galactic black hole X-ray binaries (BHBs) (comprising a
stellar mass black hole accreting matter from a partner star \citep{MillerColbert04, Fabbiano06, Roberts07}). The
X-ray spectra of ULXs can be fitted by models commonly used
for BHBs \citep{Colbert99}, and display changes in their
energy spectra similar to the state changes observed in BHBs
\citep{LaParola01}. The high luminosities are greater than that predicted by applying the Eddington
limit allowed by spherical accretion onto a typical stellar mass
black hole ($\ls 20 ~ \mathrm{M}_{\odot}$). ULXs are, however, mostly
observed outside the nucleus of their host galaxies and thus are
unlikely to be super massive black holes ($\gs 10^{5} ~ \mathrm{M}_{\odot}$), typically observed
in the centre of galaxies as active galactic nuclei (AGN). Their high
observed luminosities and the low apparent disc temperatures measured from
the energy spectra have led to suggestions that these sources
represent an until-recently unknown class of black holes with masses
between those of stellar-mass black holes and AGN around $100 - 10000
~ \mathrm{M}_{\odot}$: intermediate mass black holes (IMBHs)
\citep{Colbert99}. In contrast a number of theories have been put
forward suggesting that the high luminosities may be caused by a
stellar mass black hole emitting or accreting in different
ways. Emission processes such as anisotropic emission \citep{King01},
relativistic beaming from jets \citep{Georganopoulos02}, or accretion onto the BH over the expected
limit \citep{Begelman02} could explain some observed
features. Constraining the masses of these objects would be an
important step forward in understanding them. 

A number of different relationships between BH mass, accretion rates
and features of power spectral densities (PSDs) common to both BHBs
and AGN have already been identified
\citep{McHardy06,Remillard06,vanderKlis06,done05,Markowitz03,uttley02,belloni90}.
If these key features are seen in the PSDs of ULXs then it may provide
a way to estimate the black hole mass. 
BHBs are observed in a number of different ``states'' traditionally defined by the
dominance of components within the energy spectra, which also show
distinctly different PSDs. It is now
recognised however that the accretion rate may in fact be a more effective
parameter to discern them \citep{Remillard06}. Hardness--Intensity
diagrams following BHBs through outbursts have revealed a form
of hysteresis path that these sources follow as they move between different
states \citep{Miyamoto95}. 

A typical outburst cycle begins in a `low/hard' state in which
the X-ray spectrum is dominated by a hard
power law, with a weak contribution from a thermal component
originating in the accretion disc. The X-ray PSD shows band-limited noise that may be approximated
by a doubly broken power law or multiple broad Lorentzians
\citep{nowak00,Remillard06}. In this state the PSD often displays quasi-periodic oscillations
(QPOs). Radio emission from a steady jet is frequently observed. 
At the peak of outburst the source may reach a `very high state'
in which the X-ray spectrum shows a steep power law and a strong
thermal component. The PSD typically features a complex, broad
continuum and strong QPOs. 
The third main state is the `high/soft' state characterised by a 
dominant soft, thermal spectrum. In this state 
the PSD appears as a broken or bending power law with an
index of $-2$ above the break and $-1$ below it, which may extend to
very low frequencies \citep{reig03}. QPOs are not usually seen and the radio emission
is greatly reduced. 
Towards the end of outburst the source returns to the `low/hard' state
and the luminosity
decreases until the source reaches quiescence. 

Observations of ULXs, particularly with \xmm, have allowed for X-ray
spectral and timing analysis which appears to show some of the
features observed in BHBs. The PSDs from \xmm\ are most sensitive to
frequencies around which we would expect to see a high frequency break
(present in both the high/soft and low/hard states) if the mass of the black hole is
sufficiently large (i.e. the IMBH model predicts a break at $\sim
10^{-4} - 10^{-2}$ Hz).

Observations in all three states of BHBs, as well as all
well-constrained AGN data, show a high-frequency break above which the
PSD decays as a steep power law (index $\ls -2$). 
This break frequency may be a key diagnostic feature in establishing the properties of the system. \cite{McHardy06} showed that the break frequencies in AGN and BHBs display a correlation dependent on the mass of the black hole and its accretion rate $T_{Break} \propto {M}^{1.12}_{BH}\dot{m}^{-0.98}_{Edd}$. BHBs in the low/hard state have a second lower frequency break from a flat power law to an index of $-1$. \cite{Kording07} have altered and extended the analysis from \cite{McHardy06} to BHBs in the low/hard state through modelling the PSD as a series of Lorentzians (apparent when the PSD y-axis is plotted as $P \times f$) and then forming a relationship between accretion, BH mass and the Lorentzian relating to the high frequency break. QPOs have been detected in two ULXs, M82 X-1 and NGC 5408 X-1, indicating that the systems may have similar variability properties to BHBs
\citep{Strohmayer03,Strohmayer07}. By contrast, other studies of ULX variability
have indicated that these sources often do not show
strong short term variability \citep{Feng05,Goad06,Roberts07}.  However a break in the PSDs of ULXs NGC 5408 X-1 and M82 X-1, from a flat power law to -1 dependence, similar to that seen in the low/hard state of BHBs, has been identified below the QPOs by \cite{Soria04} and \cite{Dewangan06b,Mucciarelli06} respectively. Identifying the higher characteristic break frequency in the PSDs of ULXs could allow for a more accurate determination of their masses, and a deeper understanding of the nature of these particular sources.

 This paper describes a power spectrum analysis of the time series
 obtained from 19 \xmm\ observations of ULXs. Several of these
 observations have been analysed and discussed separately elsewhere
 (see section 2), but this paper presents the first uniform analysis
 and presentation of the power spectra from a reasonable-sized sample
 of good quality ULX observations (i.e. observations with good
 exposure time of sources with high X-ray fluxes). As noted above, some
 ULX observations show no evidence for rapid X-ray variability, but in
 almost all published studies \citep[e.g.][]{Feng05} it is not clear
 whether the lack of detected variations is due to insufficient data or
 an intrinsically low variability amplitude (by comparison with well
 studied BHBs and AGN, and also the variable ULXs). Indeed, the careful
 analysis of  Ho II X-1 by \cite{Goad06} showed the ULX to be
 intrinsically under-variable. This raises some interesting questions,
 but it is not clear whether the lack of strong variability is common
 in ULXs: in all other previous studies
 where no variability was detected, there was no attempt to constrain
 the amplitude of intrinsic variation allowed. In the
 present study we analysed the power spectra  of all 19 observations,
 and where no variability was detected, we followed the approach used
 by \cite{Goad06} to constrain the power spectrum. This allowed us to
 compare the variability properties of all the ULXs, whether variability
 could be detected or not, in a way not possible using the results of
 previously published studies.

The rest of this paper is organised as follows. Section 2 describes our sample selection criteria and search process. In Section 3 we discuss the basic data reduction techniques used and methods of temporal analysis. Section 4 explains our results in some detail including investigations into the upper limits of variation hidden within the power spectra. Finally Section 5 gives a brief discussion and conclusion.

\section{Sample Selection}
\label{subsec:sample}

We have performed a search for bright ULXs with a flux previously observed to be greater than $0.5\times10^{-12}$ erg s$^{-1}$. A sample was formed through use of sources previously identified in \cite{Stobbart06}, searching the \rosat\ catalogues of \cite{LiuBregman05} and \cite{LiuMirabel05} and by carrying out an archival study of papers studying ULXs using the ADS database (ADS\footnote{\tt{http://ads.harvard.edu/}}). A search was then made for long observations ($>$ 25 ks) of regions including these sources available by February 2008 in the $\xmm~$ public data archive (XSA\footnote{\tt{http://xmm.esac.esa.int/xsa}}) . The final sample consists of 19 observations from 16 sources with the potential to provide useful timing data (see Table \ref{table:sourcelist}). 

 As mentioned above, several of these observations have been discussed
 elsewhere in the literature, and here we briefly summarise these
 reports.  The observations of NGC 1313 X-1 and X-2, 
 NGC 5204 X-1, NGC 4945 X-2, NGC 4861 ULX and NGC 253 PSX-2 have no published
 discussion of their short-term variability.

 The observations of NGC 55 ULX were known to display interesting variability
 \citep{Stobbart05}, but this has not previously been analysed in
 detail.  No variability was detected by \cite{Feng05} in the
 observations of NGC 4395 X-1, NGC 3628 X-1, M83 ULX and NGC 2403 X-1.
 \cite{Barnard07} detected no variability in NGC 4559 X-1
 \citep[revising the previous report by][]{Cropper04}. Only the remaining
 five observations have detailed power spectral analyses published to date.
 The PSDs from the two observations of M82 X-1 have been published in
 \cite{Strohmayer03} and \cite{Dewangan06b} respectively: these show a
 strong QPO and a break in the PSD.  A similar QPO and spectral break
 has also been detected in the observation of NGC 5408 X-1
 \citep{Strohmayer07}. \cite{Dewangan06a} claimed a modestly
 significant QPO in the power spectrum of  the observation Ho IX X-1.
 Finally, as mentioned above, \cite{Goad06} presented a detailed
 analysis of the lack of variability present in the long observation of
 Ho II X-1.

\begin{table*} 

\centering
\begin{tabular}{lllrrrrr}
\hline
Source         &  Alternate Names  & Obs. Id. & RA (J2000) & Dec. (J2000) & Duration & Good Time & L$_{x}$ \\
name           &       &   &   &  & (ks) & (ks) & 10$^{39}$ erg/s \\
(1)            & (2)   & (3)   & (4) & (5)  & (6)  & (7) & (8) \\
\hline
NGC 55 ULX$^{a}$      &  & 0028740201 & 00:15:28.9 & -39 13 19.1$^{a}$ & 34.0 & 30.3 & 1.21 \\
                      &  & 0028740101 &            &             & 28.3 & 26.1 & 1.18 \\
NGC 253 PSX-2$^{b}$   & NGC 253 XMM1$^{l}$ & 0125960101 & 00:47:32.9 & -25 17 50.3$^{l}$ & 38.9 & 31.6 & 0.25 \\
NGC 1313 X-1$^{c}$    & NGC 1313 ULX1$^{j}$, IXO 7$^{k}$ & 0405090101 & 03:18:19.8 & -66 29 09.6$^{j}$ & 121.2 & 85.8  & 4.25 \\
NGC 1313 X-2$^{c}$    & NGC 1313 ULX3$^{j}$, IXO 8$^{k}$ & 0405090101 & 03:18:22.0 & -66 36 04.3$^{j}$ & 121.2 & 85.8  & 5.12 \\
NGC 2403 X-1$^{d}$    &  & 0164560901$^{r}$ & 07:36:25.9 & +65 35 38.9$^{j}$ & 77.5 & 58.3 & 2.62 \\
Holmberg II X-1$^{e}$ & PGC 23324 ULX1$^{j}$, IXO 31$^{k}$ & 0200470101$^{q}$ & 08:19:29.9 & +70 42 18.6$^{j}$ & 79.5 & 45.3 & 16.6 \\
M82 X-1$^{f}$         & Source 7$^{m}$ & 0112290201$^{i}$ & 09:55:50.2 & +69 40 47.0$^{m}$  & 27.3 & 27.4 & 21.0 \\
                      &  & 0206080101$^{u}$ &            &             & 91.0 & 57.5 & 20.5 \\
Holmberg IX X-1$^{g}$ & M81 X-9$^{n}$, PGC 28757 ULX1$^{j}$, IXO 34$^{k}$ & 0200980101$^{t}$ & 09:57:54.1 & +69 03 47.3$^{j}$ & 111.8 & 58.7 & 8.67 \\
NGC 3628 X-1$^{d}$    & IXO 39$^{k}$ & 0110980101$^{r}$ & 11:20:15.8 & +13 35 13.6$^{a}$ & 50.3 & 46.1 & 4.14 \\
NGC 4395 X-1$^{d}$    & NGC 4395 ULX1$^{j}$, IXO 53$^{k}$  & 0142830101$^{r}$ & 12:26:01.5 & +33 31 34.7$^{j}$ & 106.1 & 78.6 & 0.21 \\
NGC 4559 X-1$^{d}$    &  \parbox{1.8in}{NGC 4559 ULX2$^{j}$; ULX-7$^{p}$, IXO 65$^{k}$} & 0152170501$^{s,p}$ & 12:35:51.6 & +27 56 01.7$^{j}$ & 40.2 & 33.3  & 9.01 \\
NGC 4861 ULX$^{a}$    & NGC 4861 ULX1$^{j}$, IXO 73$^{k}$ & 0141150101 & 12:59:02.0 & +34 51 12.0$^{j}$ & 15.4 & 13.8  & 7.90 \\
NGC 4945 X-2$^{h}$    & NGC 4945 XMM1$^{l}$ & 0204870101 & 13:05:33.3 & -49 27 36.3$^{l}$ & 62.5 & 22.0 & 0.15 \\ 
NGC 5204 X-1$^{d}$    & NGC 5204 ULX1$^{j}$, IXO 77$^{k}$ & 0405690201 & 13:29:38.6 & +58 25 07.5$^{j}$ & 43.4 & 28.5 & 5.27 \\
                      &  & 0405690501 &            &             & 40.0 & 15.9 & 5.00 \\
M83 ULX$^{a}$         & NGC 5236 ULX1$^{j}$, IXO 82$^{k}$ & 0110910201$^{r}$ & 13:37:20.2 & -29 53 47.7$^{j}$ & 27.0 & 18.9 & 0.97 \\
NGC 5408 X-1$^{i}$    & NGC 5408 ULX1$^{j}$ & 0302900101$^{v}$ & 14:03:19.6 & -41 22 58.7$^{o}$ & 130.0 & 95.7 & 5.28 \\
\hline
\multicolumn{8}{c}{{\it }}\\
\end{tabular} 
\caption{Observation Details. (1) Source Name; (2) alternate names; (3) $\xmm~$ observation identifier; (4)-(5) Positions from $\xmm~$ or $\chandra~$ data; (6) $\xmm~$ observation length (ks); (7) length of good time with background flares excluded (ks); (8) Observed Luminosity measured from energy spectral fitting (see text). Footnotes indicate references as follows: Column 1-2 footnotes indicate the identification of sources with the given name(s); column 3
footnotes refer to previous power spectral analyses of the observations; column 5 footnotes provide references to source positions. $^{a}$\protect\cite{Guainazzi00}; $ ^{b}$\protect\cite{Roberts00}; $ ^{c}$\protect\cite{Stobbart06}; $ ^{d}$\protect\cite{Humphrey03}; $ ^{e}$\protect\cite{Soria06}; $ ^{f}$\protect\cite{Colbert95}; $ ^{g}$\protect\cite{Miller04}; $ ^{h}$\protect\cite{Dewangan04}; $ ^{i}$\protect\cite{Strohmayer03}; $ ^{j}$\protect\cite{LiuBregman05}; $ ^{k}$\protect\cite{Colbert02}; $ ^{l}$\protect\cite{Winter06}; $ ^{m}$\protect\cite{Matsumoto01}; $^{n}$\protect\cite{Fabbiano89}; $^{o}$\protect\cite{Kaaret03}; $^{p}$\protect\cite{Cropper04} $^{q}$\protect\cite{Goad06} $^{r}$\protect\cite{Feng05} $^{s}$\protect\cite{Barnard07} $^{t}$\protect\cite{Dewangan06a} $^{u}$\protect\cite{Dewangan06b} $^{v}$\protect\cite{Strohmayer07} }
\label{table:sourcelist}
\end{table*}

\section{Analysis}
\label{sect:analysis} 

Light-curves and energy spectra were extracted from the EPIC pn camera on \xmm. In all
observations the camera was operating in full-field mode excepting
that of Ho IX X-1 (obs Id. 0200980101) which was taken in large window
mode. The MOS cameras were not utilised due to their lower time resolution, making it harder to constrain the Poisson noise level during power spectral analysis. When the data were included they were found to add little to either the results or analysis. The data from each observation was extracted and processed using
the \xmm\ SAS version 7.1.0. Data was filtered leaving only events
with {\tt PATTERN $\le$ 4} and {\tt FLAG==0}. Events were extracted
either within the optimised radius of the source, as evaluated by
the SAS, or a radius of our own definition if other factors (such as
the source sitting close to a chip edge or near another X-ray point
source) made this unsuitable. A background light curve was taken from a rectangular region on the same chip and as close to the source as possible.

Light-curves for the 0.3-10 keV band were extracted with a time resolution of $73.4$
ms for all sources except for Ho IX X-1, the light curve of which had a time resolution of $96$ ms accounting for the different camera operational mode. Most of the ULXs are near the
aimpoint of the image and are clearly detected as separate point
sources, with a few exceptions. Most sources were therefore extracted from a region centred on the co-ordinates given in Table \ref{table:sourcelist}. NGC 4395 X-1 is close to the edge of
the chip; M82 X-1 and NGC 3628 X-1 are close to the centre of their
host galaxies and other bright X-ray sources, causing difficulties in
exactly identifying the source whilst excluding all others given the
resolution of \xmm. For M82 X-1 we followed \cite{Strohmayer03} and
extracted an 18" region centred on the co-ordinates for the point
source identified clearly in \chandra\ observations. Analysis on power
spectra taken from regions surrounding the bright ULX in the \xmm\
observations by \cite{Feng07} has suggested that this is where the QPO
originates. NGC 3628 X-1 was extracted in a 23" region centred on the
co-ordinates given in Table \ref{table:sourcelist}. The second
observation of NGC 55 ULX (obs. Id. 0028740101) is off-axis, causing a
lower average count rate than observed in the first observation
(obs. Id. 0028740201) and so the two observations were treated separately. 

In order to remove background flares a light curve was extracted from
the entire chip in the $10-15$ keV band and then criteria applied to
exclude regions where the count rate stayed continuously over three
times the mean. We only considered flares longer than 30s. Any telemetry effects where the count rate dropped to zero
for a period longer than 15 s were replaced with the mean count rate
in the source file (this had only a small effect on the overall light curve and subsequent temporal analysis). The second half of the light curve from Ho IX X-1 (obs Id. 0200980101) was particularly affected by this. The continuous regions not affected by flaring were defined as good time intervals.

Luminosities were measured through fitting each of the energy spectra with a multicoloured disc + power law model and are included in Table \ref{table:sourcelist}. 

\subsection{Timing Analysis}
\label{subsec:timana}

\begin{figure*}
\begin{center}
\hbox{
  \hspace{1.0 cm}
	\includegraphics[width=5.3cm, angle=270]{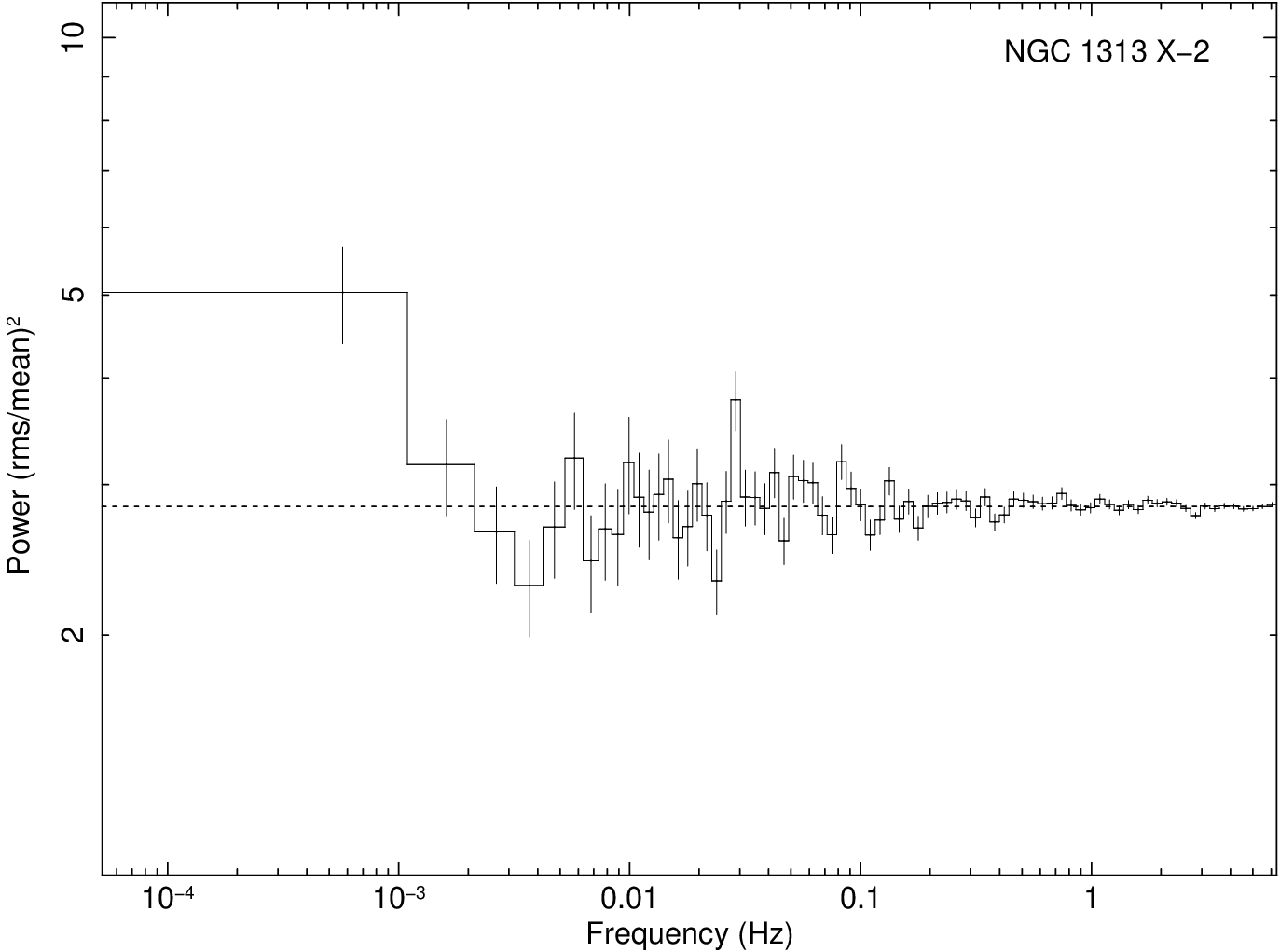} 
  \hspace{0.5 cm}
	\includegraphics[width=5.3cm, angle=270]{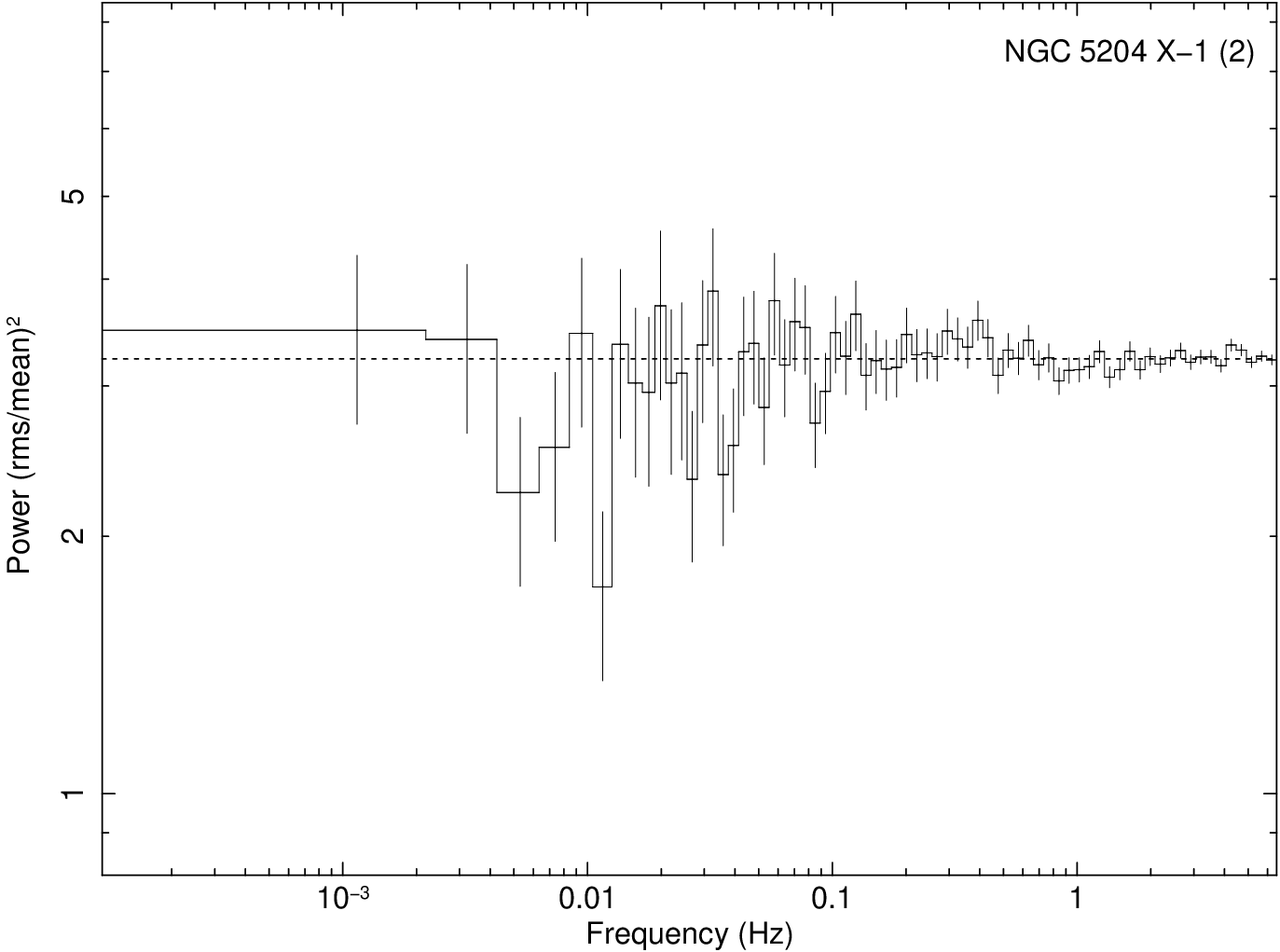} 
}
\hbox{
  \hspace{1.0 cm}
	\includegraphics[width=5.3cm, angle=270]{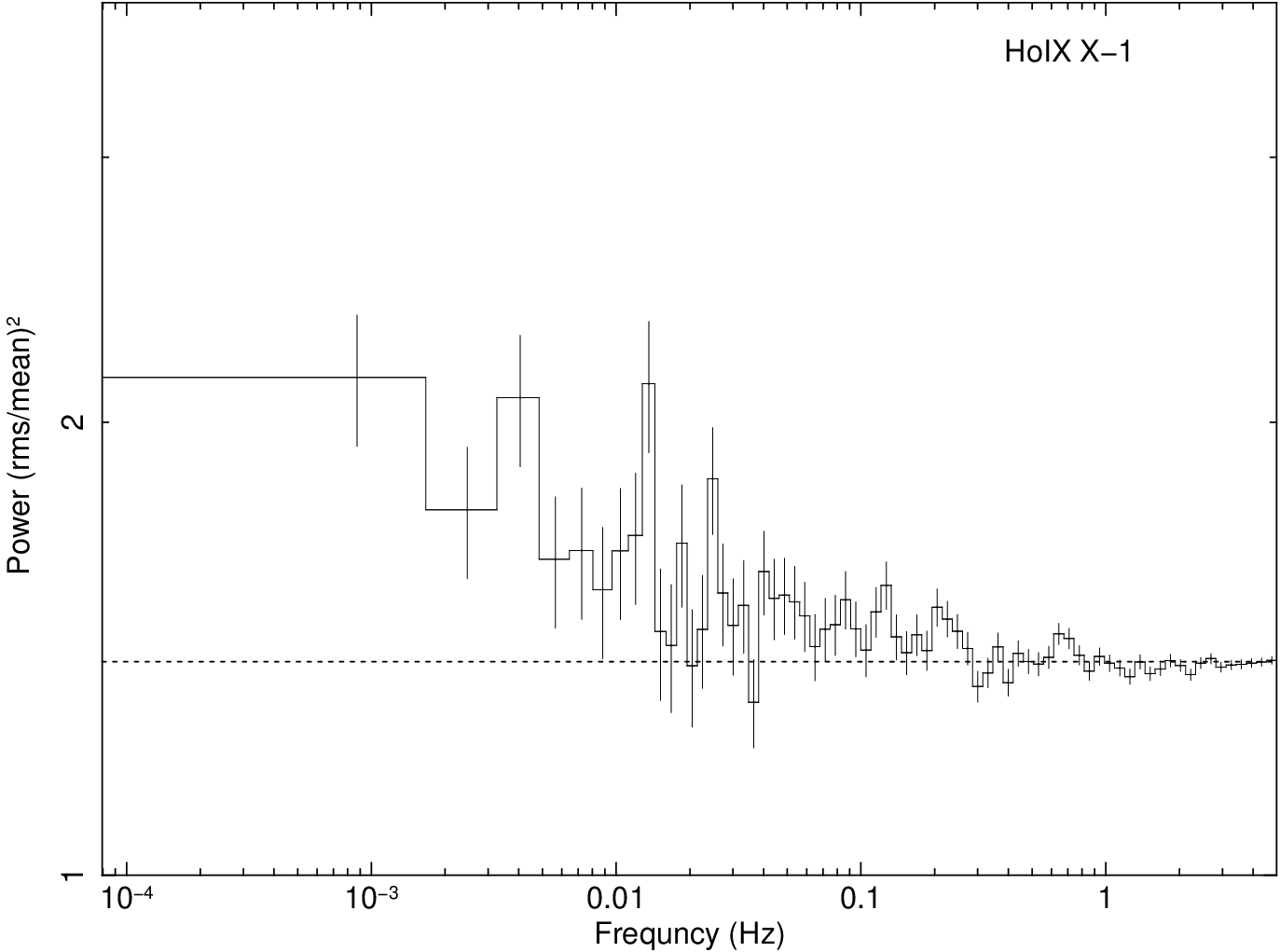} 
  \hspace{0.5 cm}
	\includegraphics[width=5.3cm, angle=270]{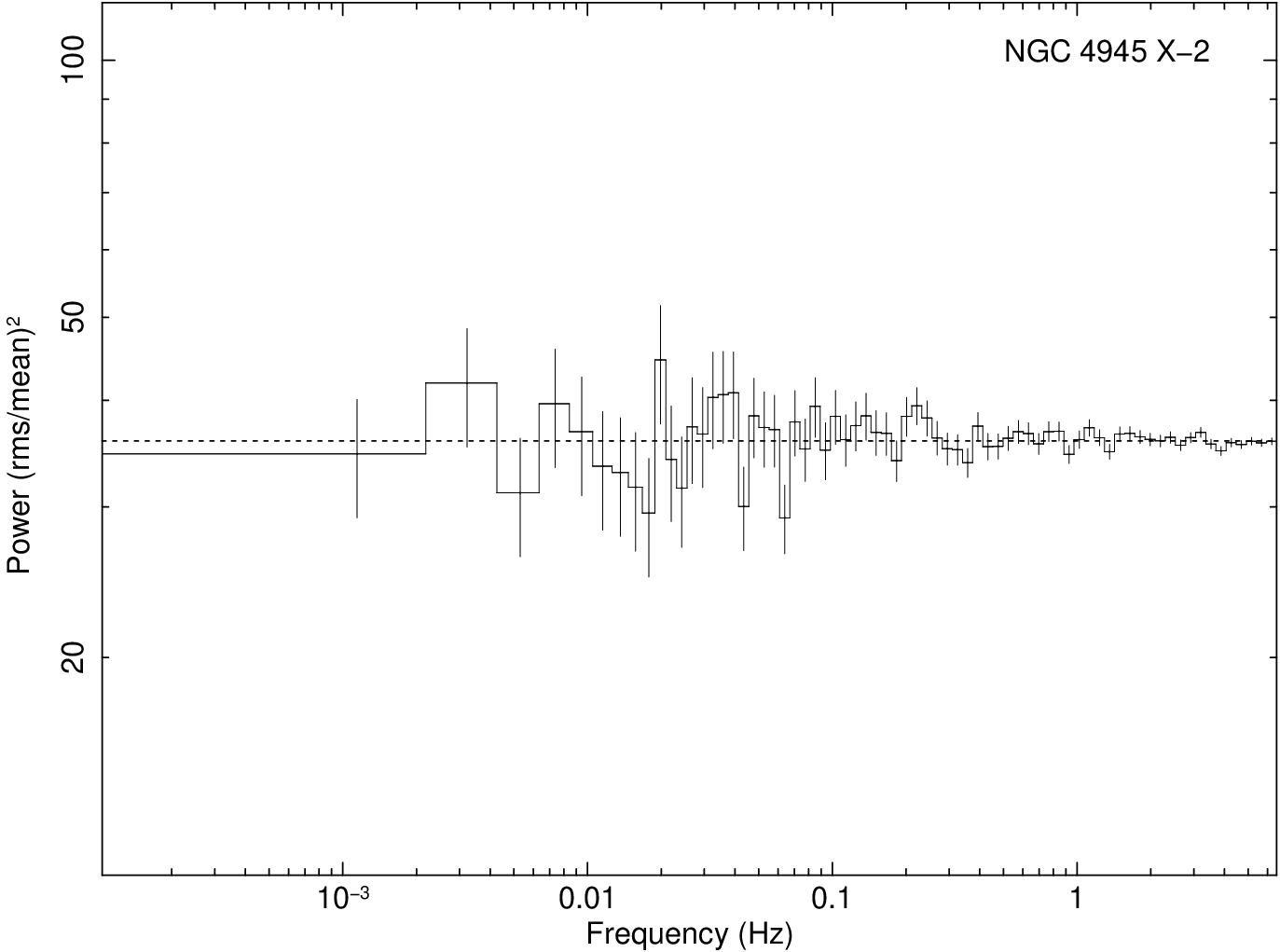} 
}
\hbox{
  \hspace{1.0 cm}
	\includegraphics[width=5.3cm, angle=270]{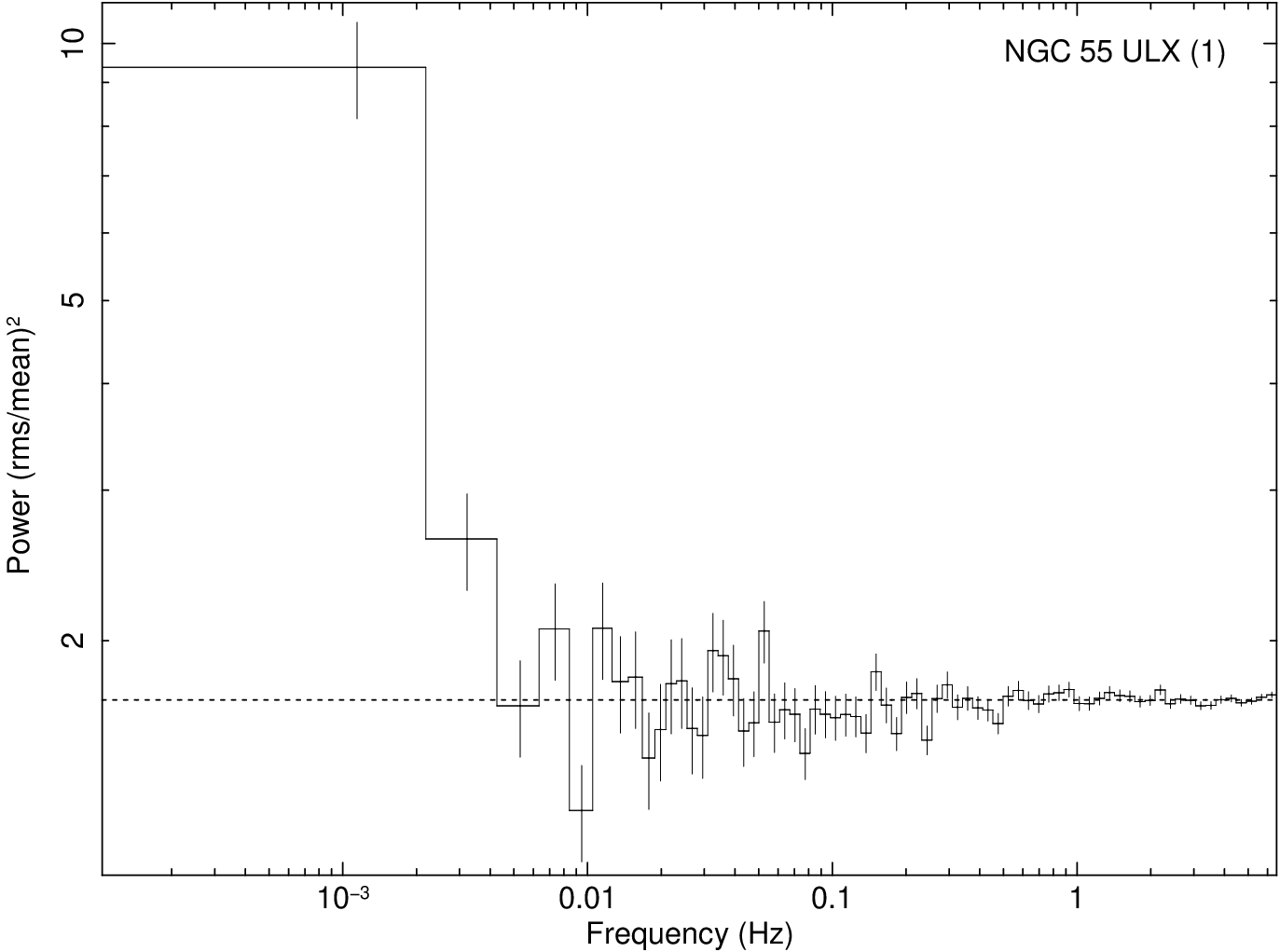} 
  \hspace{0.5 cm}
	\includegraphics[width=5.3cm, angle=270]{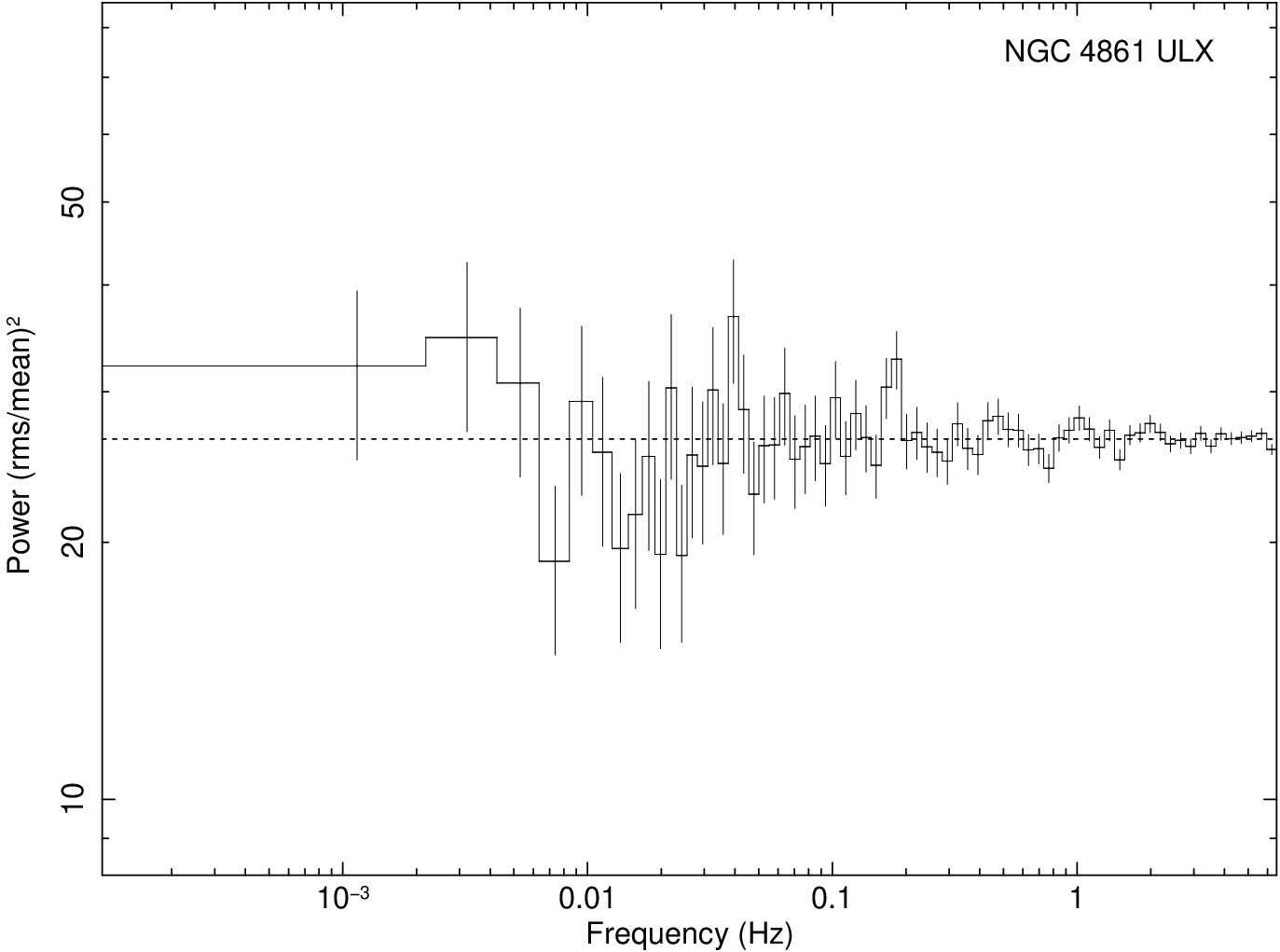} 
}
\hbox{
  \hspace{1.0 cm}
	\includegraphics[width=5.3cm, angle=270]{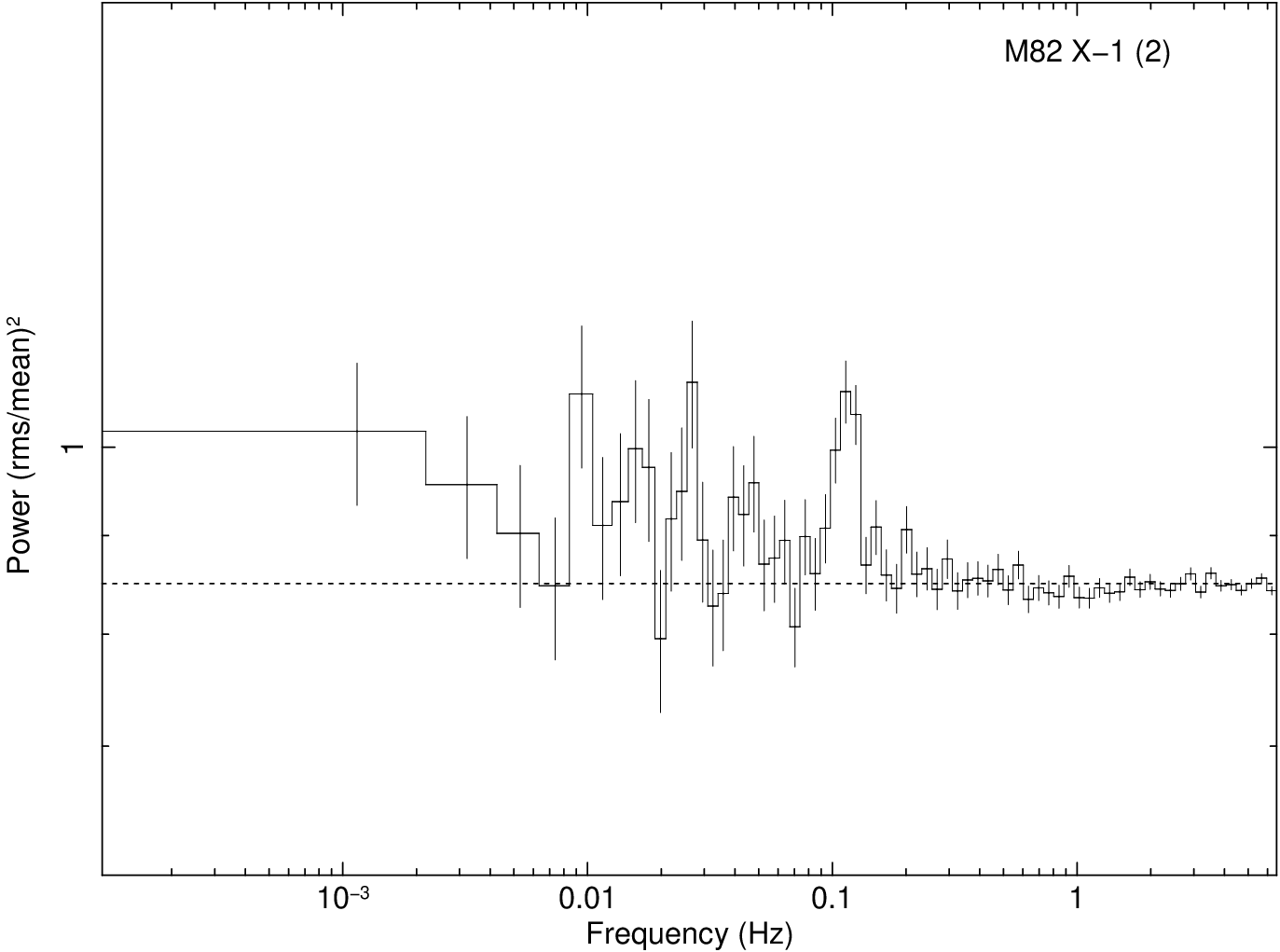} 
  \hspace{0.5 cm}
	\includegraphics[width=5.3cm, angle=270]{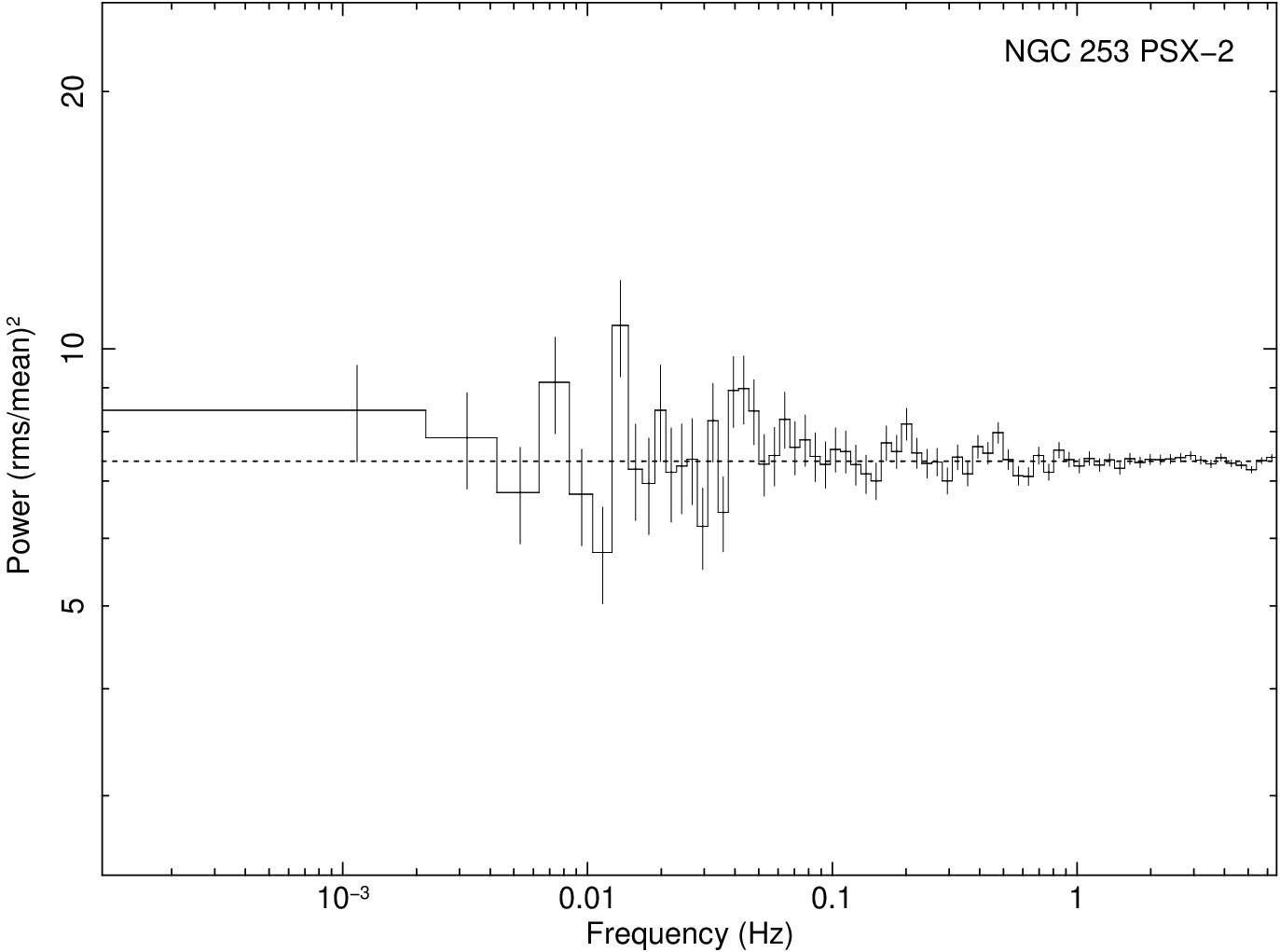}
}
\end{center}
\caption{Power spectra of 8 observations from the sample, representative of the range of variability and power seen. All light curves are split into at least 5 equal segments in time and then re-binned in frequency to give increments of 1.1f$_{0}$. The Poisson noise level for each observation is also shown (dotted line). The top six figures show comparisons between sources which display power law type variability and sources which display no strong short term variability, just Poisson noise. The figure of the second observation of M82 X-1 (obs. ID. 02060801) shows variability but not of a power law type, it is visible as excess power in the PSD above the line indicating the Poisson noise level. The QPO in M82 X-1 at 0.12 Hz is clearly visible, as is the steep, strong red-noise feature at low frequencies in the PSD from the first observation of NGC 55 ULX (obs. ID. 0028740201)}
\label{fig:PSDs}
\end{figure*}

In order to estimate the PSD from an observation the 
available good time was split into at least five segments of equal
length. The FFTs from each segment were averaged and the result
re-binned in frequency giving a factor $1.1$ increase in frequency per
bin (i.e binning over $f_{i} \rightarrow 1.1 f_{i}$), but ensuring 
there were at least $20$ points in each. 
This averaging process permits minimum-$\chi^{2}$ fitting to be used
as a maximum likelihood method \citep{vanderKlis89,papadakis93}. The
PSD was normalised to the fractional rms squared \citep{vanderKlis97}. 
Visual examination revealed no QPOs in addition to those already
reported  in M82 X-1 \citep{Strohmayer03} and NGC 5408 X-1 \citep{Strohmayer07}.   
Figure~\ref{fig:PSDs} shows a selection of eight of the sample of PSDs, these are
all plotted with the same segment size for ease of comparison, it must be noted 
that some of the PSDs used for the analysis were produced with larger segments 
and this allows a few more points of data to be visible at low frequencies.

At high frequencies the PSD is expected to be dominated by Poisson
noise, random fluctuations in the photon counts even in the absence of
intrinsic variability in the source flux. This produces a constant
(i.e. flat) additional component in the PSD \citep{vanderKlis89} with
an expected noise level (in the absence of detector non-linearities)
of approximately $2(S + B)/S^{2}$ where $S$ is the source count rate
and $B$ is the background count rate, calculated from
the counts in the good time intervals \citep{uttley02,vaughan03}. 
In order to determine the Poisson noise level in each observation 
the high
frequency region of the PSD ($\ge 0.5$ Hz) was fitted with a constant and the
normalisation compared to the expected Poisson noise level. Generally
these correspond within the $\sim 2$\% errors on the PSD
estimate. However for the two brightest sources, Ho II X-1 and M82
X-1, the measured values were $\sim 3$\% below the prediction levels.
These differences may be due to instrumental effects such as
event pile-up or dead time \cite[see also][]{Goad06}.  

\section{Results}
\label{sec:Results}

The following subsections describe the results of fitting various simple models to the power spectral data.

\subsection{Variable Sources}
\label{subsec:variasource}

In the absence of any intrinsic variability from the source the power spectrum is expected to be constant with frequency resulting from Poisson noise in the photon counting detector.
A reasonable test for the presence of intrinsic variability is therefore a $\chi^2$ goodness-of-fit test comparing the data to a constant. 
The results are given in Table \ref{tab:vars}.
Sources with a $p-$value $\leq 0.05$ were considered to show significant intrinsic variability (with this threshold one might expect $\sim 1$ false detection in a sample of $19$ observations).
In fact six sources (eight observations) showed significant variability: Ho IX X-1, NGC 5408 X-1, NGC 1313 X-1, NGC 1313 X-2, NGC 55 ULX and M82 X-1. Investigation of the PSD from NGC 4559 X-1 revealed it to be flat with no indication of the variability features analysed in \cite{Cropper04}, this is believed to be due to artificial variability caused by the analysis method as discussed in \cite{Barnard07}.

\begin{table}

\centering
\begin{tabular}{llrl}
\hline
Source         &  &  & \\
Name           & $\chi^2$/dof & P$_{null}$ & Var. \\
(1)            & (2)   & (3) & (4) \\
\hline
NGC 55 ULX      & 128.9/63 & 1.35$\times10^{-3}$ & P \\
                & 155.4/63 & $<\times10^{-6}$ & P \\
NGC 253 PSX2    & 74.8/63  & 0.255 & N \\
NGC 1313 X-1    & 110.0/63 & 3.07$\times10^{-3}$ & P* \\
NGC 1313 X-2    & 128.6/91 & 1.13$\times10^{-2}$ & P \\
NGC 2403 X-1    & 58.9/63  & 0.193 & N \\
Holmberg II X-1 & 85.3/63  & 0.554 & N \\
M82 X-1         & 108.3/63 & 3.78$\times10^{-2}$ & Pq \\
                & 160.3/63  & $<\times10^{-6}$ & Pq* \\
Holmberg IX X-1 & 261.6/63 & $<\times10^{-6}$ & P \\
NGC 3628 X-1    & 84.7/64 & 0.134 & N \\
NGC 4395 X-1    & 78.1/77 & 0.413 & N \\
NGC 4559 X-1    & 78.3/63  & 0.409 & N \\
NGC 4861 ULX    & 64.6/63 & 0.375 & N \\
NGC 4945 XMM1   & 61.1/63 & 0.746 & N \\ 
NGC 5204 X-1    & 47.2/63  & 0.875 & N \\
                & 68.8/63  & 0.276 & N \\
M83 ULX         & 158.3/63 & 0.068 & N \\
NGC 5408 X-1    & 437.9/63 & $<\times10^{-6}$ & Pq \\
\hline
\multicolumn{4}{c}{{\it }}\\

\end{tabular}
\caption{Results on fitting a constant to determine variability. (1) Source Name; (2) $\chi^{2}$ and $dof$ values for PSD after fitting with a constant model; (3) Null Hypothesis Probability for PSD fit with a constant (4) Assessment of Variability; $N$ denotes no observed variability, $P$ power law type variability, $*$ indicates the variability is of a white noise form, $q$ PSD contains a QPO}
\label{tab:vars}
\end{table}

With the exception of the second observation of M82 X-1 and the observation of NGC 1313 X-1, which display variability in the power spectrum close to the Poisson noise level \citep{Dewangan06b, Mucciarelli06}, the variable objects all show a clear rise towards low frequencies compared to the observations with flat PSDs (e.g. Ho II X-1, NGC 5204 X-1). The continuing rise seen in XTE observations of M82 X-1 with a power law of -1 \citep{Kaaret06,Kaaret07} supported by evidence of breaks seen in NGC 5408 X-1 suggests that the PSDs are best fit with a broken power law model over a broad Lorentzian. For this reason power law models were fitted to the variable sources, although the limited number of points at low frequencies made it difficult to constrain the model parameters. The results are given in Table \ref{table:spectfits}.

\begin{figure}
\begin{center}
\includegraphics[width=6.0cm, angle=270]{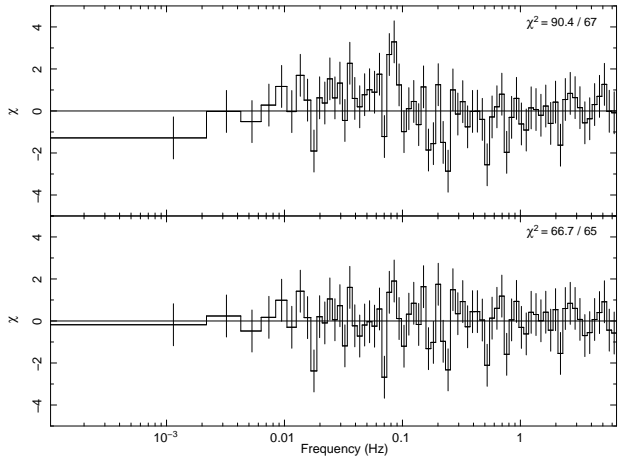}
\caption{Standardised residuals for Power law ({\it top}) and Broken Power law ({\it bottom}) fits to the PSD of NGC 1313 X-1. $~\chi^{2}$ values for the fits are also included.}
\label{fig:1313res}
\end{center}
\end{figure}

Ho IX X-1, NGC 1313 X-2, NGC 55 ULX and M82 X-1 (obs. Id. 0112290201) were fitted best with a power law (plus a constant for the Poisson noise). 
NGC 5408 X-1, M82 X-1 (obs Id 0206080101) and NGC 1313 X-1 were fitted better with a broken power law (plus a constant) as used in \cite{Strohmayer07} to model the PSD from NGC 5408 X-1. This is the first report of a broken power law PSD in NGC 1313 X-1 and in order to test the plausibility of the detection we have attempted to fit the PSD with a power law (+ constant) model, and compared the $\chi^{2}$ value from this fit to that of a broken-power law (+ constant) model which has two further free parameters. Analysing the difference between the two $\chi^{2}$ values both through direct comparison and the {\it F-test} suggest that the spectral break at 0.09 Hz is significant and necessary to the fit with $p_{null} \approx 10^{-4}$. Fig. \ref{fig:1313res} shows the standardised residuals after fitting these two models.

In the cases of M82 X-1 and NGC 5408 X-1 the obvious QPOs were modelled with narrow Lorentzians (see below).
The power law-like regions can all be fitted with slopes between $-2$ and $-0.5$. For those sources with previous analyses of their variability (M82 X-1, NGC 5408 X-1 and Ho IX X-1) the power law slopes, break frequencies and Lorentzian parameters found agree with the previous results, with the exception of the QPO identified in Holmberg IX X-1 by \cite{Dewangan06a} which was not reproduced with high significance although a continuum PSD is visible.

\begin{table*}

\centering
\begin{tabular}{lrrrrrr}
\hline
Source    & $\alpha_{1}$ & $f_{b}~$$(10^{-2}~Hz)$ & $\alpha_{2}$ & $A_{1}$ & $C_{p}$ & $\chi^{2}/dof$ \\
(1) & (2) & (3) & (4) & (5) & (6) & (7) \\
\hline
NGC 55 ULX (1)   & 1.75$\pm0.2$ & - & - & (1.53$\pm0.6$)$\times10^{-5}$ & 1.71$_{-0.02}^{+0.06}$ & 86.9/70  \\
NGC 55 ULX (2)	 & 1.96$\pm0.04$ & - & - & (6.55$\pm2.0$)$\times10^{-6}$ & 3.87$\pm0.02$ & 122.6/70  \\
NGC 1313 X-1     & -0.082$\pm0.07$ & 9.1$_{-0.9}^{+0.6}$ & 2.35$\pm1.0$ & 0.26$\pm0.06$  & 3.16$\pm0.008$ & 71.9/70 \\
NGC 1313 X-2     & 1.91$_{-0.1}^{+0.05}$ & - & - & (5.82$\pm3.0$)$\times10^{-7}$ & 3.59$\pm0.007$ & 69.3/70 \\ 
M82 X-1 (1)      & 0.70$\pm0.06$ & - & - & (1.1$\pm0.3)\times10^{-2}$ & 1.346$\pm0.006$ & 41.6/55 \\
M82 X-1 (2)      & 0.14$_{-0.05}^{+0.02}$ & 2.0$\pm1$ & 1.11$_{-0.3}^{+0.8}$ & (5.5$\pm2.2)\times10^{-2}$ & 0.847$\pm0.002$ & 66.7/70 \\
Holmberg IX X-1  & 0.48$\pm0.04$ & - & - & (2.55$\pm0.3$)$\times10^{-2}$ & 1.36$\pm0.04$ & 73.6/70  \\
NGC 5408 X-1     & -0.10$_{-0.03}^{+0.07}$ & $0.3_{-0.07}^{+0.04}$ & 1.45$_{-0.2}^{+0.5}$ & 4.0$\pm0.8$ & 2.31$\pm0.01$ & 100.2/77 \\ 
\hline
\multicolumn{7}{c}{{\it }}\\
\end{tabular}
\caption{Best fit models for the sample sources displaying variability.
	(1) Source names; 
	(2) Power law index (below break for broken power law);
 	(3) Break frequency of broken power law component in Hz; 
	(4) power law index above the break;
	(5) Normalisation of power law/broken power law component; 
	(6) Normalisation of Poisson noise component;
	(7) $\chi^2$ and $dof$ values for the fitted models. A Lorentzian component (with $Q=5$) was included
  in the models fitted to the QPOs in M82 X-1 and NGC 5408. For more detailed examination of the QPOs see \protect\cite{Strohmayer03} and \protect\cite{Strohmayer07}.}
\label{table:spectfits}
\end{table*}

\begin{figure*}
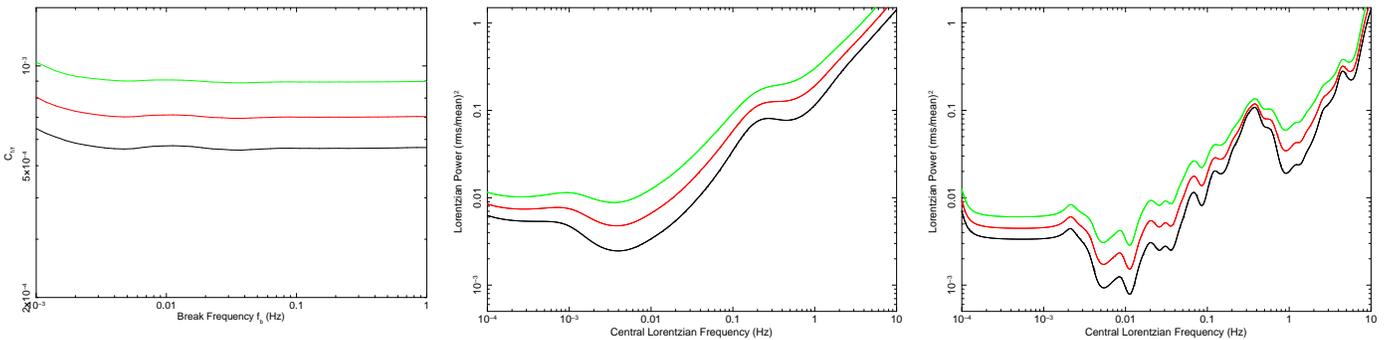


\begin{center}$
\begin{array}{lcl}
	\includegraphics[width=4.2cm, angle=270]{NGC5204_1_C_cont.epsi} &
	\includegraphics[width=4.4cm, angle=270]{ContoursNGC5204_1_0.5_final.epsi} &
	\includegraphics[width=4.4cm, angle=270]{ContoursNGC5204_1_5.0_final.epsi}
\end{array}$
\end{center}
\caption{Contour plots for our three variability models from NGC 5204 (obs Id. 0405690201), limits shown correspond to $\Delta\chi^{2} =$  1.0, 2.71 and 6.63. {\it Left:} High/soft model, broken power law -1/-2. {\it Centre:} Band-limited noise, Lorentzian (Q=0.5). {\it Right:} Narrow Lorentzian (Q=5.0)}
\label{fig:contplots}
\end{figure*}

\subsection{Establishing limits on the variability}
\label{subsect:limvaria}

The other $11$ observations show no strong evidence for intrinsic variability, despite a range of observation lengths and source fluxes, and therefore sensitivity to variability.
In these cases the variability amplitude was constrained, following the procedure of \cite{Goad06}, by assuming the intrinsic power spectrum has one of two generic forms common to the PSDs of BHBs and AGN.

The first model is a broken power law model. It is
well established that BHBs and AGN show PSDs that are usually
dominated by a broad-band noise component with a form that
resembles to first-order a broken power law. Most sources
show a frequency range in which the spectral index (PSD
slope) is approximately $-1$ -- noise with this PSD are
known as ''flicker noise'' \citep{press78}. The integral of
a $f^{-1}$ spectrum diverges (if rather slowly) to low and
high frequencies, and so must break to flatter and steeper
slopes respectively. Therefore, we may use a broken power
law as an approximate model of the generic broad-band noise
PSD.

In many AGN and ''high/soft'' state BHBs the low frequency
break (to an index flatter than $-1$) is not observed and
must occur at very low frequencies
\citep{reig03,Markowitz03,uttley02,reig02,Churazov01}. This
means a good approximation to the observed PSDs can be made
by considering a singly broken power law (BPL) with indicies
of $-1$ and $-2$ below and above some break frequency,
respectively.
\begin{equation}
P(f) = \left\{ \begin{array}{ll}
A \left(\frac{f}{f_{\rm b}}\right)^{-1} + C_P & \mbox{where
$f <f_{\rm b}$} \\
A \left(\frac{f}{f_{\rm b}}\right)^{-2} + C_P & \mbox{where
$f \ge f_{\rm b}$}
\end{array}
\right.
\end{equation}
where $A$ is a normalisation, $f_{\rm b}$ is the (high
frequency) break frequency, and $C_P$ is a constant to
account for the Poisson noise.

If we represent this spectrum in $f \times P(f)$ form the
$f^{-1}$ part of the intrinsic (i.e. after subtracting
the contribution due to Poisson noise) spectrum ($f < f_{\rm b}$) becomes
flat, with a constant level $C_{1/f} = f_{\rm b} P(f_{\rm
b}) = Af_{\rm b}$. It has been suggested that the value of
$C_{1/f}$ is fairly constant across BHBs and AGN, with
values in the range $0.005 - 0.03$ \citep{Goad06,Papadakis04}. Of course the break frequencies $f_b$ vary widely from source to source -- from 10$^{-7}$ Hz for AGN with BH masses $\sim 10^{8} ~ \mathrm{M}_{\odot}$, to $\sim 10$ Hz in stellar mass BHBs -- and on long timescales the break frequency may vary by as much as a decade in frequency for a single source \citep{belloni90,uttley02,belloni02}. For the non-varying ULX observations there was generally no observational constraint on $f_b$ and so C$_{1/f}$ was estimated (or constrained) as a function of break frequency over the range $10^{-4} - 1$ Hz. Figure \ref{fig:contplots} (left panel) shows a contour plot of the upper limit on $C_{1/f}$ as a function of $f_b$ for NGC 5204 X-1 (obs. Id. 0405690201). The contours show the value of $C_{1/f}$ at which the $\chi^2$ fit statistic increased by $1.0, 2.71$, and $6.63$ over the minimum. These correspond to the $68.3$,
$90$ and $99$ per cent quantiles of the $\chi^2$ distribution with $\nu = 1$
degrees of freedom, and are conventionally used to delimit confidence
regions on one parameter. Here they are used to define approximate limits
on the magnitude of $C_{1/f}$ as a function of $f_b$. Table \ref{tab:limits} gives $90\%$ limits on the value
of $C_{1/f}$ for each of the sources in the sample
calculated assuming two different values of $f_{\rm b}$,
namely $10^{-3}$ Hz and $1$ Hz.

For the variable sources for which an unbroken power law was
a good fit we estimate the value of $C_{1/f}$ by fitting a
BPL model. Where the observed power law index was $\approx
-2$ (i.e. in NGC 55 ULX and NGC 1313 X-2) it was assumed
that the high frequency break, $f_{\rm b}$, must lie below
the observed frequency range ($\ls few \times 10^{-3}$ Hz),
and so we assumed $f_{\rm b} = 10^{-4}$ Hz. If, on the other
hand, the observed power law index was close to $-1$ (i.e.
Holmberg IX X-1, and the first observation of M82 X-1), it
was assumed that the $f_{\rm b}$ must lie close to, or
higher than, the highest frequency at which ULX variability
can be detected above the Poisson noise. In these cases we assumed $f_{\rm b} = 1$ Hz. 

In the cases of NGC 1313 X-1, NGC 5408 X-1 and the second observation of
M82 X-1, where a break in the power law is detected, the observed indices at low
frequencies are rather low (closer to $0$ than $-1$)
indicating that the observed PSD covers the low-intermediate
part of the model. Here it is flat, breaking 
from an index of $0$ to $-1$ the low frequency flattening discussed earlier.
 In these cases the intermediate-high part
of the PSD was again assumed to break from an index of $-1$
to $-2$ at $f_{\rm b} = 1$ Hz, while the low frequency index
of $0$ was fixed, the location of the lower frequency break was left as a free
parameter.

The second model represents the band-limited noise (BLN) commonly observed in BHB low/hard states \citep{nowak00,belloni02}. In this model the variability is mostly limited to $\sim 1-2$ decades of temporal frequency  \citep{vanderKlis06}. A simple model for this type of noise is a broad (incoherent) Lorentzian:
\begin{equation}
P(f) = \frac{ R^2 Q f_0 / \pi }{ f_0^2 + Q^2 (f - f_0)^2}+ C_P
\end{equation}
where $R^2$ is a normalisation term \citep[approximately integrated power,][]{nowak00}, $f_0$
is the resonant frequency, and $Q$ is the quality factor. Again, $C_P$ is a constant to account for the Poisson noise. The $Q$-factor defines the frequency width of the Lorentzian, $Q \approx f_{0}/HWHM$ where $f_{0}$ is the central Lorentzian frequency and $HWHM$ its half width at half maximum. Typically a feature with $Q > 2$ is classified as a QPO and $Q < 2$ as band-limited noise \citep{vanderKlis89,vanderKlis06}. 
In order to model band-limited noise a value of $Q=0.5$ was assumed. The same model form but with $Q=5.0$ was used to model QPO features.
Of course, given the quality of the data the resonant frequency $f_0$ of the BLN could not be constrained, and so the strength of any BLN was estimated (or constrained) as a function of $f_0$ spanning $10^{-4} - 1$ Hz, in the same manner as with the BPL model.
In those observations that showed no obvious QPOs the same procedure was used to put limits on the strength of any QPOs as a function of QPO frequency.
The centre and right panels of Fig.~\ref{fig:contplots} show the upper limits on amplitudes of the BLN and QPO models for NGC 5204 X-1 (obs Id. 040569201). These were found through minimising the fit to the models over 1000 different values of either the central Lorentzian frequency or the break frequency over the \xmm~ frequency range. 

Table \ref{tab:limits} gives the approximate limits on the amplitudes of the two models (BLN and BPL)
assuming two representative values of the characteristic frequency in each case ($1$ mHz and $1$ Hz). The limits were calculated using a $\Delta \chi^2 = 2.71$ criterion (which corresponds to the $90$ per cent limit of a
$\chi^2$ distribution with $\nu = 1$). The fainter sources (such as NGC 4945 X-2, NGC 4395 X-1, NGC 4861 ULX, NGC  3628 X-1 and M83 ULX) have upper limits such that variability from these sources would have to be particularly strong to have been detectable. For example, in NGC 4395 X-1 a BLN spectrum could have $R \approx 20\%$ fractional rms and this would barely reach the $90\%$ confidence limit even assuming a characteristic frequency best-suited to detection ($f_0 \sim 10^{-3}$ Hz). For those sources in which variability was detected the measured values of the amplitude are given for each model. The observation of NGC 5408 X-1 gives the amplitude of a BLN spectrum (assuming $f_0 \approx 10^{-3}$ Hz) as $R \approx 18\%$ whereas the upper limit on the amplitude of the same model in the case of  NGC 5204 X-1 (obs Id. 0405690201) gives $R \ls 7\%$ even though the two observations show comparable Poisson noise levels.

\subsection{Limits on QPOs}
\label{subsect:limQPO}

The data were examined for possible QPOs by re-binning the FFT data in both frequency and time using various different bin widths.  
As discussed above a narrow ($Q = 5$) Lorentzian was used to model a QPO. For each observation the change in $\chi^2$ was calculated as a function of the two free parameters $R^{2}$ and $f_0$. An improvement of $\Delta\chi^{2} = 11.3$ was considered indicative of a QPO.
This process clearly recovered the known QPOs in NGC 5408 X-1 and M82 X-1 but did not find any further candidates. 
The known QPOs show typically $R^{2} \sim 10$\% and $f_0 \sim 2 - 5 \times 10^{-2}$ Hz.
By comparison the limits on the QPO strength derived from observations of several other bright ULXs, such as NGC 5204 X-1 and Ho II X-1, show that such QPOs would have been detected if they were present (see Fig.~\ref{fig:contplots}).
The QPO claimed in Ho IX X-1 by \cite{Dewangan06a} was not reproduced with high significance.
In these sources QPOs similar to those in NGC 5408 X-1 and M82 X-1 sources would have to be fairly muted if present.

\begin{table*}

\centering
\begin{tabular}{lrrrrr}
\hline
Source         & \multicolumn{2}{c}{Broken Power Law (BPL)} & \multicolumn{3}{c}{Band limited noise (BLN)} \\
Name           & C$_{1/f}~(10^{-4})$ & C$_{1/f}~(10^{-4})$ & R$^{2}~(10^{-2})$ & R$^{2}~(10^{-2})$ & R$_{m}^{2}~(10^{-2})$ \\
 & $10^{-3} Hz$ & $1Hz$ & $10^{-3} Hz$ & $1Hz$ & \\
(1)            & (2)   & (3) & (4) & (5)  & (6) \\
\hline
NGC 55 ULX      & - & 177$\pm46$ & - & - & 3.2$\pm0.5$ \\
                & - & 463$\pm142$ & - & - & 9.2$\pm2$ \\
NGC 253 PSX2    & $<65.4$ & $<18.5$ & $<1.64$ & $<35.2$ & - \\
NGC 1313 X-1    & - & 136$\pm46$ & - & - & 6.0$_{-2}^{+0.8}$ \\
NGC 1313 X-2    & - & 26$\pm14$ & - & - & 0.65$_{-0.1}^{+0.9}$ \\
NGC 2403 X-1    & $<2.9$ & $<2.3$ & $<0.282$ & $<15.5$ & - \\
Holmberg II X-1 & $<2.88$ & $<3.7$ & $<0.0610$ & $<2.23$ & - \\
M82 X-1         & - & 7.4$\pm5.0$ & - & - & 4.2$\pm2$ \\
                & - & 22.1$\pm6.0$ & - & - & 2.5$\pm0.5$ \\
Holmberg IX X-1 & - & 6.0$\pm$1.7 & - & - & 0.88$\pm0.2$ \\
NGC 3628 X-1    & $<170$ & $<43.1$ & $<4.49$ & $<33.1$ & - \\
NGC 4395 X-1    & $<260$ & $<64.7$ & $<5.58$ & $<92.8$ & - \\
NGC 4559 X-1    & $<7.11$ & $<28.4$ & $<1.72$ & $<43.4$ & - \\
NGC 4861 ULX    & $<401$ & $<99$ & $<10.8$ & $<206$ & - \\
NGC 4945 XMM1   & $<180$ & $<51.8$ & $<5.37$ & $<293$ & - \\ 
NGC 5204 X-1    & $<21$ & $<5.9$ & $<0.494$ & $<10.5$ & - \\
                & $<34.3$ & $<5.4$ & $<0.539$ & $<13.3$ & - \\
M83 ULX         & $<29.1$ & $<64$ & $<7.56$ & $<42.3$ & - \\
NGC 5408 X-1    & - & 104$\pm11$ & - & - & 2.99$\pm0.5$ \\
\hline
\multicolumn{6}{c}{{\it }}\\

\end{tabular}
\caption{Upper limits on the fitted model amplitudes using two different power spectral models. The estimate of C$_{1/f}$ from the BPL model at f$_{b} = 10^{-3}$ Hz for Ho II X-1 differs from the value found in \protect\cite{Goad06} differs due to the use of different binning for the fitting process. Columns: (1) Source Name; (2) C$_{1/f}$ from the broken power law model with f$_{b} = 10^{-3}$ Hz; (3) For non-varying sources: C$_{1/f}$ from the broken power model with the f$_{b} = 1Hz$; For varying sources: C$_{1/f}$ from fit to model according to shape of PSD: See text; (4) Upper limit on fractional rms within a Lorentzian of coherence Q=0.5 fit at $10^{-3}$ Hz; (5) Upper limit on fractional rms within a Lorentzian of coherence Q=0.5 fit at 1 Hz; (6) Fit normalisation values to Band limited noise when present. All limits correspond to $\Delta\chi^{2} = 2.71$ (which is approximately 90$\%$ confidence).}
\label{tab:limits}
\end{table*}

 
\section{Discussion}
\label{sect:disco} 

The main results of the temporal analysis of $19$ observations of $16$ sources may be summarised as follows:

\begin{itemize}

\item
Six sources (eight observations) showed significant intrinsic flux variability in the $10^{-4} - 1$ Hz range (NGC 5408 X-1, NGC 1313 X-1, NGC 1313 X-2, NGC 55 ULX, M82 X-1 and Ho IX X-1).

\item
Of these six, two show strong QPO features (NGC 5408 X-1 and M82 X-1, as previously identified) and all show a continuum spectrum consistent with a power law or broken power law model.

\item
Of the remaining $10$ sources, the intrinsic variability amplitudes were constrained by comparing the data to simple PSD models based on BHB observations.

\item
The limits obtained for several of the observations (e.g. NGC 4559 X-1, NGC 5204 X-1, Ho II X-1) are such that the strength of the intrinsic variability, in the frequency range observed, must be substantially lower than the observed power seen in the variable sources.

\end{itemize}

\subsection{Break Frequencies and BH mass}
\label{subsec:breakfreq}

It is in principle possible to draw inferences about the nature of the putative accreting black
hole in the ULXs if analogies can be formed between their observed
properties and those of BHBs and AGN. Perhaps the simplest and best studied
relation linking BHBs and AGN is the correlation between the PSD break
frequencies and BH mass (e.g. \cite{Markowitz03}; \cite{done05}; \cite{McHardy06}), 
a correlation which may well include ULXs.

The 3 PSDs of NGC 5408 X-1, M82 X-1 (obs. ID. 0206080101) and NGC 1313 X-1 show a break from an index
of $\sim -1$ above the break to $\sim 0$ below, which resembles the `low
frequency break' observed in low/hard BHB observations
\citep{belloni90,nowak00,Remillard06}. \cite{Soria04} have used the break frequency in NGC 5408 X-1 to derive an
expected BH mass of $\sim 100 ~M_{\odot}$. For M82 X-1 \cite{Dewangan06b}
measured the PSD break frequency in relation to the QPO frequency and drew
a comparison with BHB features in order to estimate the BH mass at $25-520
~M_{\odot}$. \cite{Casella08} applied the relationship in \cite{Kording07},
which extends the analysis from \cite{McHardy06} into systems in a low/hard
state, to estimate a BH mass between $115-1305 ~ M_{\odot}$ for NGC 5408 X-1
and $95-1260 ~ M_{\odot}$ for M82 X-1.

 By contrast the break in NGC 1313 X-1 at 0.09 Hz is comparable to those observed in BHBs, for example the observed break of Cyg X-1 in the low/hard state is 0.08 Hz. Cyg X-1 is thought to have a BH mass of $\sim 10 M_{\odot}$. However the observed luminosity of Cyg X-1 in this state is only $\sim 4 \times 10^{37}$ erg/s \citep{Nowak99} a factor of 100 lower. Either NGC 1313 X-1 hosts a commensurately
larger BH mass than Cyg X-1, or it does not but is accreting at a much
higher relative rate ($L/L_{\rm Edd} > 1$), or at least appears to be due to
anisotropic emission (or some combination of these). In either case the flat
PSD below $0.09$ Hz is difficult to account for. The flat slope is most commonly observed in 'low/hard'
state BHBs, but this state is thought to occur when
$L/L_{\rm Edd} \ls$ 0.05 \citep{Remillard06}. Assuming NGC
1313 X-1 to be in such a state would therefore require
$M_{\rm BH} \gs 600 \mathrm{M}_{\odot}$, yet the frequency
of the break matches well the expectation for a $M_{\rm BH}
\sim 10 \mathrm{M}_{\odot}$ black hole. This supports the
suggestion that ULXs may represent a different,
'ultraluminous' state \citep{Roberts07}.

NGC 1313 X-2, NGC 55 ULX and Ho IX X-1 show excess power rising to low frequencies
below $\sim 5 \times 10^{-3}$ Hz. As can be seen in Figure \ref{fig:PSDs},
the PSD from the first observation of NGC 55 ULX (obs. ID. 0028740201)
shows a steep red noise region at low frequencies, also visible in the
PSD from the second observation (obs. ID. 0028740101). When the two observations are fitted with a
power law + constant model, spectral indices of 1.75$\pm$0.2 and 1.96$\pm$0.4
are found respectively, with no sign of a low frequency break. This would
suggest, assuming the system is in the high/soft state, that the high frequency
break would have to be below 10$^{-4}$ Hz, indicative of a BH with a mass in
excess of $10^{6}~M_{\odot}$ \citep{Roberts07} more typical of AGN.
However it must be pointed out that dips in the light-curve have been identified
\citep{Stobbart05} and that these features affect the PSD on timescales commensurate with the strong variability power seen in its PSD. The power laws seen in the PSDs from Ho IX X-1 and NGC 1313 X-2 have spectral indices of $\sim$0.5 and $\sim$1.91 respectively. The few rising points visible in the PSD of NGC 1313 X-2 do not allow for accurate fitting and so little can be derived from this value. For Ho IX X-1 however, the slope is
 much clearer and we have carried out a basic analysis. We have been able to 
 constrain the likelihood of a brake to a slope of -2 to be above 0.1 Hz. This would
 indicate a BH with a mass below 1000 M$_{\odot}$ \citep{Roberts07}.

The possibility of estimating the masses of ULXs by
comparing characteristic (e.g. break or QPO) frequencies
with those observed in BHBs and AGN, is dependent on there being a
suitably strong analogy mapping PSD characteristics between
the source, together with a simple mass-frequency scaling
relation \citep{McHardy06}. The recent discovery of an apparent high frequency QPO
within a AGN \citep{Gierlinski08} indicates that these features are common across the range of
observed sources. If however ULXs do not display variability in forms
typical to either BHBs or AGN and instead exist in a new `Ultraluminous state' \citep{Roberts07} with distinct and different characteristics to those states observed in AGN and BHBs, then these mass estimates will no longer be
reliable.

\begin{figure*}
\includegraphics[width=6.5cm,angle=90]{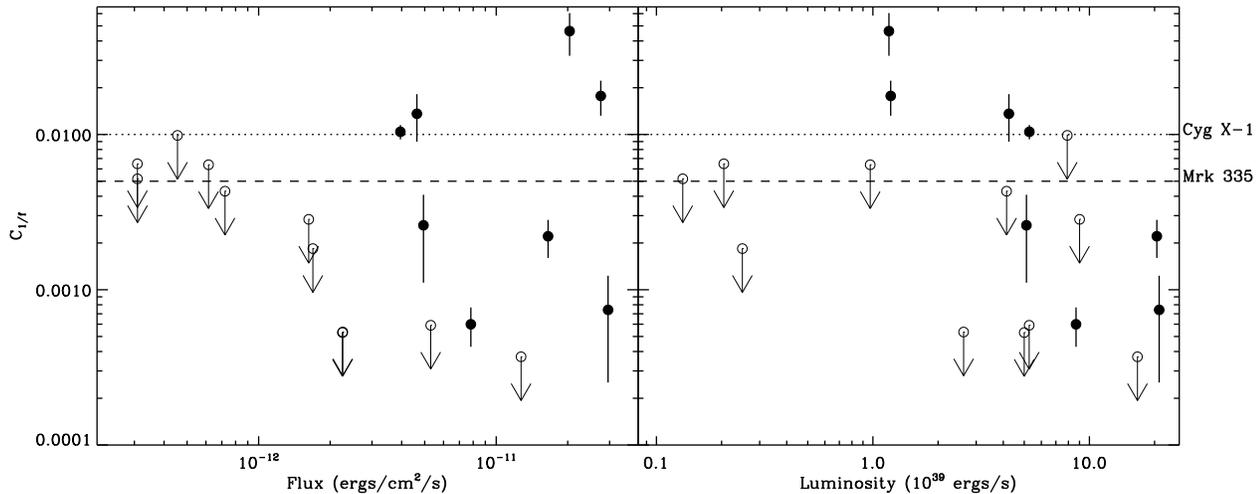}
\caption{Upper limits to high/soft model against Flux {\it(left)} and luminosity {\it(right)} for f$_{b}$ = 1 mHz. Points defined as varying are represented by filled circles, whilst non-varying sources are unfilled. Also shown are the measured values of C$_{1/f}$ for BHB Cygnus X-1 \protect\citep{Uttley07} and the AGN Mrk 335 \protect\citep{Arevalo08}, for reference the luminosities of these sources are $\sim$3$\times$10$^{37}$ erg/s and $\sim$3$\times$10$^{43}$ erg/s respectively.}
\label{fig:lumfl}
\end{figure*}

\subsection{Variability vs Flux and Luminosity}
\label{subsec:fllum}
The two main properties of an observation that determine the sensitivity to
intrinsic spectral power at a given frequency are the length of the
observation $T$ (the number of `cycles' of that frequency that were
observed) and the count rate of the source (which determines the Poisson
noise in the observation). Denoting the mean count rate as $I$ we may
approximate the significance (in units of $\sigma$) of an excess in power
over the Poisson noise level by $n_{\sigma}(f) \approx \frac{1}{2} I P_S(f)
\sqrt{T \times \Delta f(f)}$ \citep[see][]{vanderKlis89b} where $P_S(f)$ is the
power density due to the intrinsic variations in the source flux, and
$\Delta f(f)$ gives the frequency binning. Thus, with similar observation
lengths $T$ and frequency re-binning $\Delta f(f)$ the sensitivity should
scale as $n_{\sigma} \propto I P_S(f)$ and so if a detection is made when
$n_{\sigma}$ exceeds some number the limit on the power that can be detected
scales as $P_S(f) \propto I^{-1}$.

Fig.~\ref{fig:lumfl} (left panel) shows the estimated variability
amplitudes (detections and upper limits) against source flux. Also on these plots are typical values for C$_{1/f}$ observed in a BHB Cyg X-1 and the AGN Mrk 335, illustrating the consistency in C$_{1/f}$ between BHBs and AGN. The upper limits clearly improve (i.e. give tighter limits) at higher fluxes, as
expected. In particular, they indicate it is in general not possible to
detect even strong variability ($C_{1/f} > 0.05$) in sources fainter than $\sim
10^{-12}$ erg s$^{-1}$ cm$^{-1}$. Fig.~\ref{fig:lumfl} (right panel) shows
the same variability amplitudes against observed X-ray luminosity, with no
strong trend apparent. Above $10^{39}$ erg s$^{-1}$ (i.e. definite ULX
luminosities) there is a range of $\sim 30$ in the variability amplitude,
confounding any efforts to identify source states on the basis of
luminosity-variability correlations. Considering only those source for which variability was detected (filled circles) there appears to be a correlation suggesting a decrease in variability with increasing luminosity. However there are two reasons why this could be misleading. The first is that the two highest estimated values of C$_{1/f}$ are the observations of NGC 55 ULX. This source shows dipping episodes in its variability \citep{Stobbart05}, not seen in other ULXs, and this
anomalous behaviour enhances the variability amplitude. Removing these points substantially decreases the apparent significance of any luminosity-amplitude anti-correlation. Secondly,
 including the limits from observations that did not have variability detections
 shows that some lower luminosity sources ($\sim 10^{39}$ erg s$^{-1}$) must
 have very low amplitude variability, opposite to the prediction of a general
 luminosity-amplitude anti-correlation. When these points are considered there appears to be no obvious correlations within the plot.
 
\subsection{Are some ULXs significantly less variable?}
\label{subsec:novaria}

\begin{figure}
\includegraphics[width=6.0cm,angle=270]{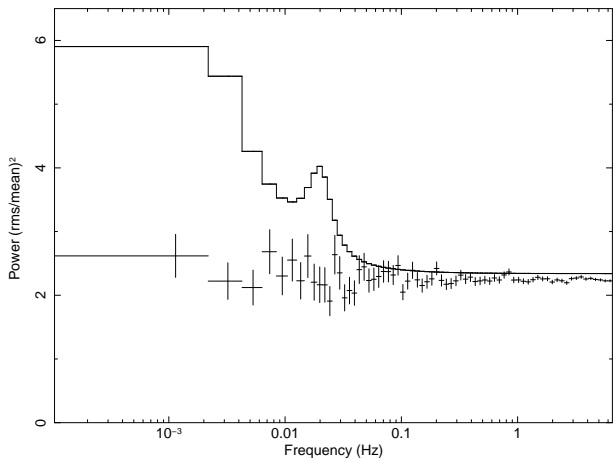}
\caption{Average power spectra of NGC 5204 (obs Id. 0405690501) with the best fitting
model for NGC 5408 X-1, demonstrating clear absence of variability in the former source
even though the observed luminosities from the two observations are very similar.}
\label{fig:varier}
\end{figure}

The upper limit contours from four bright sources (Ho II X-1, NGC 2403
X-1, NGC 5204 X-1 and NGC 4559 X-1) show that these sources are significantly less
variable than others (such as NGC 55 ULX, NGC 5408 X-1, etc.). Fig.~\ref{fig:varier}
compares the data from NGC 5204 X-1 with the model obtained by fitting the data from NGC
5408 X-1 (these two have similar luminosities and count rates, and hence Poisson noise levels in their
PSDs). 
The figure clearly shows the variability seen in NGC 5408 X-1 to be
absent or greatly suppressed in NGC 5204 X-1. 

The difference between the strength of variability observed in many ULXs and
that expected by analogy with BHBs was confirmed using simulated
data. Two PSDs of Cyg X-1 -- one showing BLN ("low/hard state") and one
showing BPL ("high/soft state")\footnote{The observations were taken from
the {\it RXTE} archive: the BLN data came from proposal P10236 made on 1996
December 16-18 and the BPL data came from P10512 made on taken 1996 June
18.} -- were used as examples of BHB variability. Random time series were
generated from these PSDs \citep[using the method of][]{Timmer95} and given
the same mean count rate and exposure time as the \xmm\ observation of
Holmberg II X-1. Poisson noise was included at the appropriate level, and the
simulated data were analysed in the same manner as the real ULX data.
Variability was clearly detected (i.e. a constant was rejected by a $\chi^{2}$
goodness-of-fit test) in the simulated data for both PSD shapes. The same was true when the
PSD frequencies were reduced (by factors of up to $10^5$) prior to generating the time
series - i.e. simulating the effects of larger
black hole masses by scaling the variability timescales.

The results confirm that if Holmberg II X-1 was variable with a PSD
shape and amplitude similar to that seen in Cyg X-1 its variability should
have been detected (at high significance) in the \xmm\ observations. This
is true whether the PSD shape is BPL or BLN, and holds when the characteristic
variability frequencies are reduced as might be expected for IMBHs. 
The results of the simulations confirm the the PSD fitting
results (discussed above), namely that the apparent lack of variability in 
the observations of sources such as Holmberg II X-1
and NGC 5204 X-1 must be intrinsic to the sources, not a result of
insufficient observations.

\subsection{Interpreting the lack of variability}

There are two fundamentally different ways to account for the lack of
variability in the ULX observations: as an observational effect caused
by the limited energy or frequency bandpass of the observations, or as
a result of the sources being intrinsically under-variable.
We consider some specific possibilities below.

The energy spectra of BHBs in their higher accretion rate states are
often dominated by a (quasi-)thermal component associated with the
accretion disc. By comparison with the non-thermal (power law-like)
hard X-ray continuum, this spectral component shows little short term
variability \citep{Churazov01}. LMC X-3 is an example of a BHB that
often shows very low amplitude short term variability in the ``high/soft''
state \citep[dominated by thermal disc emission;][]{nowak2001}. 
See also a discussion of this point in the context of BHB-AGN
comparisons in \citet{done05}. It
could be that a similar component 
dominates the spectra of the under-variable ULXs within the \xmm\
bandpass.
If this was the case we would expect the thermal emission to peak
around $\sim 1-2$ keV, higher than expected for IMBH models, which predict a much cooler
temperature $\sim 0.2$ keV \citep{Miller03, Miller04, Stobbart06},
i.e. this solution requires BH masses similar to Galactic BHBs. 
One problem with this hypothesis is that the energy
spectra of ULXs often appear more complex than the thermal-dominated
spectra of "high/soft" BHBs when viewed in the highest quality XMM-Newton
observations (\cite{Stobbart06, Gladstone09}), although
even then there are some ULXs that do appear to have modified disc-like
spectra (e.g. \cite{Vierdayanti06, Hui08}).
A prediction of this hypothesis is that the sub-variable ULXs should
show stronger variability at higher energies, where the spectrum
is dominated by variable non-thermal emission.

\cite{Goad06} suggested that the lack of variability seen in Holmberg
II X-1 may be
related to the so-called `$\chi$-state' observed in GRS 1915+105, where
the variability power is
concentrated in a BLN spectrum at high frequencies ($> 0.1$ Hz) and
there is often little variability on longer
timescales \citep{Belloni00}. 
The integrated power over the $10^{-4} - 0.1$ Hz range is typically
only $\sim 10^{-3}$ \citep[fractional variance; see e.g.][]{Zdziarski05} which is close to the detection
limits of the ULX observations. 
Again, this explanation requires BH masses comparable to that of GRS
1915+105 \citep[$\sim 14$ M$_{\odot}$;][]{greiner2001} as a
significantly larger BH mass would shift the BLN variability into the frequency
bandpass of the ULX observations ($10^{-4} - 0.1$ Hz) where it would
be detectable.
A prediction of this hypothesis is that ULXs should show significant
variability power at higher frequencies (e.g. $0.2 - 20$ Hz), in the form of
a BLN component and possibly strong QPOs, as does GRS 1915+105.

The above explanations hypothesise that the variability in ULXs is
similar to that of BHBs (and AGN) but is often not observed due to the
limited energy and/or frequency bandpass of the available data.
Another possibility is that the X-ray emission from ULXs is
intrinsically less variable (on frequencies $>10^{-3}$ Hz) than
expected by analogy with other accreting systems.

The timescales involved are sufficiently large to rule out at least
one model for the suppression of variability within the ULX system:
scattering in an extended region. 
Recent studies of the X-ray spectra of ULXs
\citep[e.g.][]{Done06, Stobbart06, Goad06} have put forward the idea of
an optically thick corona of hot electrons that modifies the emerging energy
spectrum. 
The timescale for the diffusion of photons through an electron scattering
medium is of order $ \Delta t \sim R \tau_{es} / c$, where $R$ is the
size of the scattering region and $\tau_{es}$ ($\gtrsim 1$) is the optical depth to
electron scattering \citep[see e.g.][and references therein]{Nowak99,
Miyamoto88}. On timescales shorter than $\Delta t$ the variability will be
strongly attenuated.
The ULX observations show weak or absent variability on
timescales at least as large as $\sim 10^2$ s, which would require $R \tau_{es} \sim
100$ lt-s.
This is implausibly large; even with $M_{\rm BH} \sim 100$ M$_{\odot}$
an optical depth of $\tau_{es} \sim 10 - 20$ \citep[see][]{Gladstone09} would need a scattering region
as large as $R \sim 2 \times 10^{9}$ m ($\sim 7 \times 10^3$ gravitational radii), comparable to the
binary separation of Cyg X-1. This analysis suggests electron
scattering is not a viable mechanism for suppressing variability in
ULXs on such long timescales, although this is still a promising model for the energy spectra which require a much smaller emission region than needed here.

Of course, it remains plausible that ULXs are examples of an accretion
state that is simply not found (or very rare) among the commonly
observed BHBs and AGN, and which displays very stable X-ray emission.
Nevertheless, any mechanism for explaining the lack of variability in some
ULXs faces a challenge: the X-ray spectra of the variable (e.g. NGC
5408 X-1) and non-variable (NGC 5204 X-1) ULXs are quite similar 
\citep{Soria04, Roberts06, Stobbart06} while their
variability properties appear very different (see Fig.~\ref{fig:varier}). This apparent inconsistency requires further study to understand fully, multiple observations of a few ULXs could allow for further clarification of this problem.

Longer observations with \xmm\ would allow the frequency range to be extended to lower
frequencies and could reveal in more detail the PSDs of ULXs such as
NGC 55 ULX and NGC 1313 X-2, allowing better comparisons with BHBs and AGN.
Observations of ULXs with more powerful X-ray missions (such as {\it Astro-H} and the
proposed {\it International X-ray Observatory}) are needed before 
ULX variability can be studied at the higher frequencies and in the higher energy bands
needed to test the hypotheses discussed above.


\section*{acknowledgements}

We thank an anonymous referee for a 
careful reading of our manuscript and a number of helpful
suggestions.
LMH acknowledges support from an STFC studentship.
This research has made use of NASA's Astrophysics Data System.
This paper is based on observations obtained with \xmm, an ESA science mission with
instruments and contributions directly funded by ESA Member States and
the USA (NASA).
This research has made use of the NASA/IPAC Extragalactic Database
(NED) which is operated by the Jet Propulsion Laboratory, California
Institute of Technology, under contract with the National Aeronautics
and Space Administration.


\bibliographystyle{mn2e}
\bibliography{ULXsample}

\begin{thebibliography}{}

\bibitem[\protect\citeauthoryear{{Ar{\'e}valo}, {McHardy} \&
  {Summons}}{{Ar{\'e}valo} et~al.}{2008}]{Arevalo08}
{Ar{\'e}valo} P.,  {McHardy} I.~M.,    {Summons} D.~P.,  2008, \mnras, 388, 211

\bibitem[\protect\citeauthoryear{{Barnard}, {Trudolyubov}, {Kolb}, {Haswell},
  {Osborne} \& {Priedhorsky}}{{Barnard} et~al.}{2007}]{Barnard07}
{Barnard} R.,  {Trudolyubov} S.,  {Kolb} U.~C.,  {Haswell} C.~A.,  {Osborne}
  J.~P.,    {Priedhorsky} W.~C.,  2007, \aap, 469, 875

\bibitem[\protect\citeauthoryear{{Begelman}}{{Begelman}}{2002}]{Begelman02}
{Begelman} M.~C.,  2002, \apjl, 568, L97

\bibitem[\protect\citeauthoryear{{Belloni} \& {Hasinger}}{{Belloni} \&
  {Hasinger}}{1990}]{belloni90}
{Belloni} T.,  {Hasinger} G.,  1990, \aap, 227, L33

\bibitem[\protect\citeauthoryear{{Belloni}, {Klein-Wolt}, {M{\'e}ndez}, {van
  der Klis} \& {van Paradijs}}{{Belloni} et~al.}{2000}]{Belloni00}
{Belloni} T.,  {Klein-Wolt} M.,  {M{\'e}ndez} M.,  {van der Klis} M.,    {van
  Paradijs} J.,  2000, \aap, 355, 271

\bibitem[\protect\citeauthoryear{{Belloni}, {Psaltis} \& {van der
  Klis}}{{Belloni} et~al.}{2002}]{belloni02}
{Belloni} T.,  {Psaltis} D.,    {van der Klis} M.,  2002, \apj, 572, 392

\bibitem[\protect\citeauthoryear{{Casella}, {Ponti}, {Patruno}, {Belloni},
  {Miniutti} \& {Zampieri}}{{Casella} et~al.}{2008}]{Casella08}
{Casella} P.,  {Ponti} G.,  {Patruno} A.,  {Belloni} T.,  {Miniutti} G.,
  {Zampieri} L.,  2008, \mnras, p.~666

\bibitem[\protect\citeauthoryear{{Churazov}, {Gilfanov} \&
  {Revnivtsev}}{{Churazov} et~al.}{2001}]{Churazov01}
{Churazov} E.,  {Gilfanov} M.,    {Revnivtsev} M.,  2001, \mnras, 321, 759

\bibitem[\protect\citeauthoryear{{Colbert} \& {Mushotzky}}{{Colbert} \&
  {Mushotzky}}{1999}]{Colbert99}
{Colbert} E.~J.~M.,  {Mushotzky} R.~F.,  1999, \apj, 519, 89

\bibitem[\protect\citeauthoryear{{Colbert}, {Petre}, {Schlegel} \&
  {Ryder}}{{Colbert} et~al.}{1995}]{Colbert95}
{Colbert} E.~J.~M.,  {Petre} R.,  {Schlegel} E.~M.,    {Ryder} S.~D.,  1995,
  \apj, 446, 177

\bibitem[\protect\citeauthoryear{{Colbert} \& {Ptak}}{{Colbert} \&
  {Ptak}}{2002}]{Colbert02}
{Colbert} E.~J.~M.,  {Ptak} A.~F.,  2002, \apjs, 143, 25

\bibitem[\protect\citeauthoryear{{Cropper}, {Soria}, {Mushotzky}, {Wu},
  {Markwardt} \& {Pakull}}{{Cropper} et~al.}{2004}]{Cropper04}
{Cropper} M.,  {Soria} R.,  {Mushotzky} R.~F.,  {Wu} K.,  {Markwardt} C.~B.,
  {Pakull} M.,  2004, \mnras, 349, 39

\bibitem[\protect\citeauthoryear{{Dewangan}, {Griffiths} \& {Rao}}{{Dewangan}
  et~al.}{2006}]{Dewangan06a}
{Dewangan} G.~C.,  {Griffiths} R.~E.,    {Rao} A.~R.,  2006, \apjl, 641, L125

\bibitem[\protect\citeauthoryear{{Dewangan}, {Miyaji}, {Griffiths} \&
  {Lehmann}}{{Dewangan} et~al.}{2004}]{Dewangan04}
{Dewangan} G.~C.,  {Miyaji} T.,  {Griffiths} R.~E.,    {Lehmann} I.,  2004,
  \apjl, 608, L57

\bibitem[\protect\citeauthoryear{{Dewangan}, {Titarchuk} \&
  {Griffiths}}{{Dewangan} et~al.}{2006}]{Dewangan06b}
{Dewangan} G.~C.,  {Titarchuk} L.,    {Griffiths} R.~E.,  2006, \apjl, 637, L21

\bibitem[\protect\citeauthoryear{{Done} \& {Gierli{\'n}ski}}{{Done} \&
  {Gierli{\'n}ski}}{2005}]{done05}
{Done} C.,  {Gierli{\'n}ski} M.,  2005, \mnras, 364, 208

\bibitem[\protect\citeauthoryear{{Done} \& {Kubota}}{{Done} \&
  {Kubota}}{2006}]{Done06}
{Done} C.,  {Kubota} A.,  2006, \mnras, 371, 1216

\bibitem[\protect\citeauthoryear{{Fabbiano}}{{Fabbiano}}{1989}]{Fabbiano89}
{Fabbiano} G.,  1989, \araa, 27, 87

\bibitem[\protect\citeauthoryear{{Fabbiano}}{{Fabbiano}}{2006}]{Fabbiano06}
{Fabbiano} G.,  2006, Advances in Space Research, 38, 2937

\bibitem[\protect\citeauthoryear{{Feng} \& {Kaaret}}{{Feng} \&
  {Kaaret}}{2005}]{Feng05}
{Feng} H.,  {Kaaret} P.,  2005, \apj, 633, 1052

\bibitem[\protect\citeauthoryear{{Feng} \& {Kaaret}}{{Feng} \&
  {Kaaret}}{2007}]{Feng07}
{Feng} H.,  {Kaaret} P.,  2007, \apj, 668, 941

\bibitem[\protect\citeauthoryear{{Georganopoulos}, {Aharonian} \&
  {Kirk}}{{Georganopoulos} et~al.}{2002}]{Georganopoulos02}
{Georganopoulos} M.,  {Aharonian} F.~A.,    {Kirk} J.~G.,  2002, \aap, 388, L25

\bibitem[\protect\citeauthoryear{{Gierli{\'n}ski}, {Middleton}, {Ward} \&
  {Done}}{{Gierli{\'n}ski} et~al.}{2008}]{Gierlinski08}
{Gierli{\'n}ski} M.,  {Middleton} M.,  {Ward} M.,    {Done} C.,  2008, \nat,
  455, 369

\bibitem[\protect\citeauthoryear{{Gladstone}, {Roberts} \& {Done}}{{Gladstone}
  et~al.}{2009}]{Gladstone09}
{Gladstone} J.~C.,  {Roberts} T.,    {Done} C.,  2009, \mnras, submitted

\bibitem[\protect\citeauthoryear{{Goad}, {Roberts}, {Reeves} \&
  {Uttley}}{{Goad} et~al.}{2006}]{Goad06}
{Goad} M.~R.,  {Roberts} T.~P.,  {Reeves} J.~N.,    {Uttley} P.,  2006, \mnras,
  365, 191

\bibitem[\protect\citeauthoryear{{Greiner}, {Cuby} \& {McCaughrean}}{{Greiner}
  et~al.}{2001}]{greiner2001}
{Greiner} J.,  {Cuby} J.~G.,    {McCaughrean} M.~J.,  2001, Nature, 414, 522

\bibitem[\protect\citeauthoryear{{Guainazzi}, {Matt}, {Brandt}, {Antonelli},
  {Barr} \& {Bassani}}{{Guainazzi} et~al.}{2000}]{Guainazzi00}
{Guainazzi} M.,  {Matt} G.,  {Brandt} W.~N.,  {Antonelli} L.~A.,  {Barr} P.,
  {Bassani} L.,  2000, \aap, 356, 463

\bibitem[\protect\citeauthoryear{{Hui} \& {Krolik}}{{Hui} \&
  {Krolik}}{2008}]{Hui08}
{Hui} Y.,  {Krolik} J.~H.,  2008, \apj, 679, 1405

\bibitem[\protect\citeauthoryear{{Humphrey}, {Fabbiano}, {Elvis}, {Church} \&
  {Balucinska-Church}}{{Humphrey} et~al.}{2003}]{Humphrey03}
{Humphrey} P.~J.,  {Fabbiano} G.,  {Elvis} M.,  {Church} M.~J.,
  {Balucinska-Church} M.,  2003, VizieR Online Data Catalog, 734, 40134

\bibitem[\protect\citeauthoryear{{Kaaret}, {Corbel}, {Prestwich} \&
  {Zezas}}{{Kaaret} et~al.}{2003}]{Kaaret03}
{Kaaret} P.,  {Corbel} S.,  {Prestwich} A.~H.,    {Zezas} A.,  2003, Science,
  299, 365

\bibitem[\protect\citeauthoryear{{Kaaret} \& {Feng}}{{Kaaret} \&
  {Feng}}{2007}]{Kaaret07}
{Kaaret} P.,  {Feng} H.,  2007, \apj, 669, 106

\bibitem[\protect\citeauthoryear{{Kaaret}, {Simet} \& {Lang}}{{Kaaret}
  et~al.}{2006}]{Kaaret06}
{Kaaret} P.,  {Simet} M.~G.,    {Lang} C.~C.,  2006, \apj, 646, 174

\bibitem[\protect\citeauthoryear{{King}, {Davies}, {Ward}, {Fabbiano} \&
  {Elvis}}{{King} et~al.}{2001}]{King01}
{King} A.~R.,  {Davies} M.~B.,  {Ward} M.~J.,  {Fabbiano} G.,    {Elvis} M.,
  2001, \apjl, 552, L109

\bibitem[\protect\citeauthoryear{{K{\"o}rding}, {Migliari}, {Fender},
  {Belloni}, {Knigge} \& {McHardy}}{{K{\"o}rding} et~al.}{2007}]{Kording07}
{K{\"o}rding} E.~G.,  {Migliari} S.,  {Fender} R.,  {Belloni} T.,  {Knigge} C.,
     {McHardy} I.,  2007, \mnras, 380, 301

\bibitem[\protect\citeauthoryear{{La Parola}, {Peres}, {Fabbiano}, {Kim} \&
  {Bocchino}}{{La Parola} et~al.}{2001}]{LaParola01}
{La Parola} V.,  {Peres} G.,  {Fabbiano} G.,  {Kim} D.~W.,    {Bocchino} F.,
  2001, \apj, 556, 47

\bibitem[\protect\citeauthoryear{{Liu} \& {Bregman}}{{Liu} \&
  {Bregman}}{2005}]{LiuBregman05}
{Liu} J.-F.,  {Bregman} J.~N.,  2005, \apjs, 157, 59

\bibitem[\protect\citeauthoryear{{Liu} \& {Mirabel}}{{Liu} \&
  {Mirabel}}{2005}]{LiuMirabel05}
{Liu} Q.~Z.,  {Mirabel} I.~F.,  2005, \aap, 429, 1125

\bibitem[\protect\citeauthoryear{{Markowitz}, {Edelson}, {Vaughan}, {Uttley},
  {George}, {Griffiths}, {Kaspi}, {Lawrence}, {McHardy}, {Nandra}, {Pounds},
  {Reeves}, {Schurch} \& {Warwick}}{{Markowitz} et~al.}{2003}]{Markowitz03}
{Markowitz} A.,  {Edelson} R.,  {Vaughan} S.,  {Uttley} P.,  {George} I.~M.,
  {Griffiths} R.~E.,  {Kaspi} S.,  {Lawrence} A.,  {McHardy} I.,  {Nandra} K.,
  {Pounds} K.,  {Reeves} J.,  {Schurch} N.,    {Warwick} R.,  2003, \apj, 593,
  96

\bibitem[\protect\citeauthoryear{{Matsumoto}, {Tsuru}, {Koyama}, {Awaki},
  {Canizares}, {Kawai}, {Matsushita} \& {Kawabe}}{{Matsumoto}
  et~al.}{2001}]{Matsumoto01}
{Matsumoto} H.,  {Tsuru} T.~G.,  {Koyama} K.,  {Awaki} H.,  {Canizares} C.~R.,
  {Kawai} N.,  {Matsushita} S.,    {Kawabe} R.,  2001, \apjl, 547, L25

\bibitem[\protect\citeauthoryear{{McClintock} \& {Remillard}}{{McClintock} \&
  {Remillard}}{2006}]{Remillard06}
{McClintock} J.~E.,  {Remillard} R.~A.,  2006, {Black hole binaries}.
Compact stellar X-ray sources, pp 157--213

\bibitem[\protect\citeauthoryear{{McHardy}, {Koerding}, {Knigge}, {Uttley} \&
  {Fender}}{{McHardy} et~al.}{2006}]{McHardy06}
{McHardy} I.~M.,  {Koerding} E.,  {Knigge} C.,  {Uttley} P.,    {Fender} R.~P.,
   2006, \nat, 444, 730

\bibitem[\protect\citeauthoryear{{Miller}, {Fabbiano}, {Miller} \&
  {Fabian}}{{Miller} et~al.}{2003}]{Miller03}
{Miller} J.~M.,  {Fabbiano} G.,  {Miller} M.~C.,    {Fabian} A.~C.,  2003,
  \apjl, 585, L37

\bibitem[\protect\citeauthoryear{{Miller}, {Fabian} \& {Miller}}{{Miller}
  et~al.}{2004}]{Miller04}
{Miller} J.~M.,  {Fabian} A.~C.,    {Miller} M.~C.,  2004, \apjl, 614, L117

\bibitem[\protect\citeauthoryear{{Miller} \& {Colbert}}{{Miller} \&
  {Colbert}}{2004}]{MillerColbert04}
{Miller} M.~C.,  {Colbert} E.~J.~M.,  2004, International Journal of Modern
  Physics D, 13, 1

\bibitem[\protect\citeauthoryear{{Miyamoto}, {Kitamoto}, {Hayashida} \&
  {Egoshi}}{{Miyamoto} et~al.}{1995}]{Miyamoto95}
{Miyamoto} S.,  {Kitamoto} S.,  {Hayashida} K.,    {Egoshi} W.,  1995, \apjl,
  442, L13

\bibitem[\protect\citeauthoryear{{Miyamoto}, {Kitamoto}, {Mitsuda} \&
  {Dotani}}{{Miyamoto} et~al.}{1988}]{Miyamoto88}
{Miyamoto} S.,  {Kitamoto} S.,  {Mitsuda} K.,    {Dotani} T.,  1988, \nat, 336,
  450

\bibitem[\protect\citeauthoryear{{Mucciarelli}, {Casella}, {Belloni},
  {Zampieri} \& {Ranalli}}{{Mucciarelli} et~al.}{2006}]{Mucciarelli06}
{Mucciarelli} P.,  {Casella} P.,  {Belloni} T.,  {Zampieri} L.,    {Ranalli}
  P.,  2006, \mnras, 365, 1123

\bibitem[\protect\citeauthoryear{{Nowak}}{{Nowak}}{2000}]{nowak00}
{Nowak} M.~A.,  2000, \mnras, 318, 361

\bibitem[\protect\citeauthoryear{{Nowak}, {Vaughan}, {Wilms}, {Dove} \&
  {Begelman}}{{Nowak} et~al.}{1999}]{Nowak99}
{Nowak} M.~A.,  {Vaughan} B.~A.,  {Wilms} J.,  {Dove} J.~B.,    {Begelman}
  M.~C.,  1999, \apj, 510, 874

\bibitem[\protect\citeauthoryear{{Nowak}, {Wilms}, {Heindl}, {Pottschmidt},
  {Dove} \& {Begelman}}{{Nowak} et~al.}{2001}]{nowak2001}
{Nowak} M.~A.,  {Wilms} J.,  {Heindl} W.~A.,  {Pottschmidt} K.,  {Dove} J.~B.,
    {Begelman} M.~C.,  2001, MNRAS, 320, 316

\bibitem[\protect\citeauthoryear{{Okada}, {Dotani}, {Makishima}, {Mitsuda} \&
  {Mihara}}{{Okada} et~al.}{1998}]{Okada98}
{Okada} K.,  {Dotani} T.,  {Makishima} K.,  {Mitsuda} K.,    {Mihara} T.,
  1998, \pasj, 50, 25

\bibitem[\protect\citeauthoryear{{Papadakis}}{{Papadakis}}{2004}]{Papadakis04}
{Papadakis} I.~E.,  2004, \mnras, 348, 207

\bibitem[\protect\citeauthoryear{{Papadakis} \& {Lawrence}}{{Papadakis} \&
  {Lawrence}}{1993}]{papadakis93}
{Papadakis} I.~E.,  {Lawrence} A.,  1993, \mnras, 261, 612

\bibitem[\protect\citeauthoryear{{Press}}{{Press}}{1978}]{press78}
{Press} W.~H.,  1978, Comments on Astrophysics, 7, 103

\bibitem[\protect\citeauthoryear{{Reig}, {Papadakis} \& {Kylafis}}{{Reig}
  et~al.}{2002}]{reig02}
{Reig} P.,  {Papadakis} I.,    {Kylafis} N.~D.,  2002, A\&A, 383, 202

\bibitem[\protect\citeauthoryear{{Reig}, {Papadakis} \& {Kylafis}}{{Reig}
  et~al.}{2003}]{reig03}
{Reig} P.,  {Papadakis} I.,    {Kylafis} N.~D.,  2003, \aap, 398, 1103

\bibitem[\protect\citeauthoryear{{Roberts}}{{Roberts}}{2007}]{Roberts07}
{Roberts} T.~P.,  2007, \apss, 311, 203

\bibitem[\protect\citeauthoryear{{Roberts}, {Kilgard}, {Warwick}, {Goad} \&
  {Ward}}{{Roberts} et~al.}{2006}]{Roberts06}
{Roberts} T.~P.,  {Kilgard} R.~E.,  {Warwick} R.~S.,  {Goad} M.~R.,    {Ward}
  M.~J.,  2006, \mnras, 371, 1877

\bibitem[\protect\citeauthoryear{{Roberts} \& {Warwick}}{{Roberts} \&
  {Warwick}}{2000}]{Roberts00}
{Roberts} T.~P.,  {Warwick} R.~S.,  2000, \mnras, 315, 98

\bibitem[\protect\citeauthoryear{{Soria}, {Fender}, {Hannikainen}, {Read} \&
  {Stevens}}{{Soria} et~al.}{2006}]{Soria06}
{Soria} R.,  {Fender} R.~P.,  {Hannikainen} D.~C.,  {Read} A.~M.,    {Stevens}
  I.~R.,  2006, \mnras, 368, 1527

\bibitem[\protect\citeauthoryear{{Soria}, {Motch}, {Read} \& {Stevens}}{{Soria}
  et~al.}{2004}]{Soria04}
{Soria} R.,  {Motch} C.,  {Read} A.~M.,    {Stevens} I.~R.,  2004, \aap, 423,
  955

\bibitem[\protect\citeauthoryear{{Stobbart}, {Roberts} \& {Warwick}}{{Stobbart}
  et~al.}{2004}]{Stobbart05}
{Stobbart} A.-M.,  {Roberts} T.~P.,    {Warwick} R.~S.,  2004, \mnras, 351,
  1063

\bibitem[\protect\citeauthoryear{{Stobbart}, {Roberts} \& {Wilms}}{{Stobbart}
  et~al.}{2006}]{Stobbart06}
{Stobbart} A.-M.,  {Roberts} T.~P.,    {Wilms} J.,  2006, \mnras, 368, 397

\bibitem[\protect\citeauthoryear{{Strohmayer} \& {Mushotzky}}{{Strohmayer} \&
  {Mushotzky}}{2003}]{Strohmayer03}
{Strohmayer} T.~E.,  {Mushotzky} R.~F.,  2003, \apjl, 586, L61

\bibitem[\protect\citeauthoryear{{Strohmayer}, {Mushotzky}, {Winter}, {Soria},
  {Uttley} \& {Cropper}}{{Strohmayer} et~al.}{2007}]{Strohmayer07}
{Strohmayer} T.~E.,  {Mushotzky} R.~F.,  {Winter} L.,  {Soria} R.,  {Uttley}
  P.,    {Cropper} M.,  2007, \apj, 660, 580

\bibitem[\protect\citeauthoryear{{Timmer} \& {K\"onig}}{{Timmer} \&
  {K\"onig}}{1995}]{Timmer95}
{Timmer} J.,  {K\"onig} M.,  1995, A\&A, 300, 707

\bibitem[\protect\citeauthoryear{{Uttley}}{{Uttley}}{2007}]{Uttley07}
{Uttley} P.,  2007, in {Ho} L.~C.,  {Wang} J.-W.,  eds, The Central Engine of
  Active Galactic Nuclei Vol.~373 of Astronomical Society of the Pacific
  Conference Series, {X-ray Variability of Radio-quiet AGN}.
p.~149

\bibitem[\protect\citeauthoryear{{Uttley}, {McHardy} \& {Papadakis}}{{Uttley}
  et~al.}{2002}]{uttley02}
{Uttley} P.,  {McHardy} I.~M.,    {Papadakis} I.~E.,  2002, \mnras, 332, 231

\bibitem[\protect\citeauthoryear{{van der Klis}}{{van der
  Klis}}{1989a}]{vanderKlis89}
{van der Klis} M.,  1989a, in {Phillips} K.~J.~H.,  ed., Thermal-Non-Thermal
  Interactions in Solar Flares {Fourier techniques in X-ray timing.}.
pp 27--69

\bibitem[\protect\citeauthoryear{{van der Klis}}{{van der
  Klis}}{1989b}]{vanderKlis89b}
{van der Klis} M.,  1989b, \araa, 27, 517

\bibitem[\protect\citeauthoryear{{van der Klis}}{{van der
  Klis}}{1997}]{vanderKlis97}
{van der Klis} M.,  1997, in {Babu} G.~J.,  {Feigelson} E.~D.,  eds,
  Statistical Challenges in Modern Astronomy II {Quantifying Rapid Variability
  in Accreting Compact Objects}.
pp 321--+

\bibitem[\protect\citeauthoryear{{van der Klis}}{{van der
  Klis}}{2006}]{vanderKlis06}
{van der Klis} M.,  2006, {Rapid X-ray Variability}.
Compact stellar X-ray sources, pp 39--112

\bibitem[\protect\citeauthoryear{{Vaughan}, {Edelson}, {Warwick} \&
  {Uttley}}{{Vaughan} et~al.}{2003}]{vaughan03}
{Vaughan} S.,  {Edelson} R.,  {Warwick} R.~S.,    {Uttley} P.,  2003, \mnras,
  345, 1271

\bibitem[\protect\citeauthoryear{{Vierdayanti}, {Mineshige}, {Ebisawa} \&
  {Kawaguchi}}{{Vierdayanti} et~al.}{2006}]{Vierdayanti06}
{Vierdayanti} K.,  {Mineshige} S.,  {Ebisawa} K.,    {Kawaguchi} T.,  2006,
  \pasj, 58, 915

\bibitem[\protect\citeauthoryear{{Winter}, {Mushotzky} \& {Reynolds}}{{Winter}
  et~al.}{2006}]{Winter06}
{Winter} L.~M.,  {Mushotzky} R.~F.,    {Reynolds} C.~S.,  2006, \apj, 649, 730

\bibitem[\protect\citeauthoryear{{Zdziarski}, {Gierli{\'n}ski}, {Rao},
  {Vadawale} \& {Miko{\l}ajewska}}{{Zdziarski} et~al.}{2005}]{Zdziarski05}
{Zdziarski} A.~A.,  {Gierli{\'n}ski} M.,  {Rao} A.~R.,  {Vadawale} S.~V.,
  {Miko{\l}ajewska} J.,  2005, \mnras, 360, 825

\end{thebibliography}

\bsp

\label{lastpage}

\end{document}